\documentclass[longauth]{aa}
\usepackage{txfonts}
\usepackage{natbib}
\usepackage{natbib}
\bibpunct{(}{)}{;}{a}{}{,} 
\usepackage{longtable}
\usepackage{multirow}
\usepackage{color}
\usepackage{graphicx}

\newcommand{\lya}{\ifmmode {\rm Ly}\alpha \else Ly$\alpha$\fi}
\usepackage{graphicx}
\usepackage{txfonts}
\usepackage{stackengine}
\usepackage{afterpage}
\usepackage{rotating}
\DeclareUnicodeCharacter{2010}{-}
\begin{document}

\title{The VANDELS ESO public spectroscopic survey:  observations and first data release}
\author{L. Pentericci\inst{1}
\and R. J. McLure\inst{2}  
\and B. Garilli\inst{3} 
\and O. Cucciati\inst{4}
\and P. Franzetti\inst{3}
\and A. Iovino\inst{5}
\and R. Amorin \inst{6,7}
\and M. Bolzonella\inst{4}
\and A. Bongiorno\inst{1}
\and A. C. Carnall\inst{2}
\and M. Castellano\inst{1}
\and A. Cimatti\inst{8,24}
\and M. Cirasuolo\inst{9}
\and F. Cullen\inst{2}
\and S. DeBarros\inst{10}
\and J. S. Dunlop\inst{2}
\and D. Elbaz\inst{11}
\and S. Finkelstein\inst{12}
\and A. Fontana\inst{1}
\and F. Fontanot\inst{13} 
\and M. Fumana\inst{3}
\and A. Gargiulo\inst{3}
\and L. Guaita\inst{1,14}
\and W. Hartley\inst{15}
\and M. Jarvis\inst{16}
\and S. Juneau\inst{11}
\and W. Karman\inst{17}
\and D. Maccagni\inst{3}
\and F. Marchi\inst{1}
\and E. Marmol-Queralto\inst{2}
\and K. Nandra\inst{18,19}
\and E. Pompei\inst{20}
\and L. Pozzetti\inst{4}
\and M. Scodeggio\inst{3}
\and V. Sommariva\inst{8}
\and M. Talia\inst{4,8}
\and O. Almaini\inst{21}
\and I. Balestra\inst{22}
\and S. Bardelli\inst{4}
\and E. F. Bell\inst{23}
\and N. Bourne\inst{2}
\and R.A.A. Bowler\inst{16}
\and M. Brusa\inst{8}
\and F. Buitrago\inst{24,25}
\and C. Caputi\inst{17}
\and P. Cassata\inst{26}
\and S. Charlot\inst{27}
\and A. Citro\inst{8}
\and G. Cresci\inst{28}
\and S. Cristiani\inst{13}
\and E. Curtis-Lake\inst{27}
\and M. Dickinson\inst{29}
\and S.M. Faber\inst{30}
\and G. Fazio\inst{31}
\and H.C. Ferguson\inst{32}
\and F. Fiore\inst{1}
\and M. Franco\inst{11}
\and J.P.U. Fynbo\inst{33}
\and A. Galametz\inst{18}
\and A. Georgakakis\inst{18}
\and M. Giavalisco\inst{34}
\and A. Grazian\inst{1}
\and N.P. Hathi\inst{32}
\and I. Jung\inst{12}
\and S. Kim\inst{35}
\and A. M. Koekemoer\inst{32}
\and Y. Khusanova\inst{36}
\and O. Le F\`evre\inst{36}
\and J. Lotz\inst{32}
\and F. Mannucci\inst{28}
\and D. Maltby\inst{21}
\and K. Matsuoka\inst{28}
\and D. McLeod\inst{2}
\and H. Mendez-Hernandez\inst{26}
\and J. Mendez-Abreu\inst{37,38}
\and M. Mignoli\inst{4}
\and M. Moresco\inst{4,8}
\and A. Mortlock\inst{2}
\and M. Nonino\inst{13}
\and M. Pannella\inst{39}
\and C. Papovich\inst{12}
\and P. Popesso\inst{40}
\and D.J. Rosario\inst{41}
\and P. Rosati\inst{42}
\and M. Salvato\inst{18,40}
\and P. Santini\inst{1}
\and D. Schaerer\inst{10}
\and C. Schreiber\inst{43}
\and D. Stark\inst{44}
\and L.A.M. Tasca\inst{36}
\and R. Thomas\inst{20}
\and T. Treu\inst{45}
\and E. Vanzella\inst{4}
\and V. Wild\inst{46}
\and C. Williams\inst{44}
\and G. Zamorani \inst{4}
\and E. Zucca\inst{4}
}
\institute{
INAF-Osservatorio AStronomico di Roma , via Frascati 33, I-00078 Monteporzio Catone, Italy 
\and
Institute for Astronomy, University of Edinburgh, Royal Observatory, Edinburgh, EH9 3HJ, UK)
\and
INAF-Istituto di Astrofisica Spaziale e Fisica Cosmica Milano, via Bassini 15, I-20133, Milano, Italy
\and
INAF-Osservatorio di Astrofisica e Scienza dello Spazio di Bologna, via Gobetti 93/3, I-40129, Bologna, Italy
\and 
INAF-Osservatorio Astronomico di Brera, via Brera 28, I-20122 Milano, Italy
\and 
Kavli Institute for Cosmology, University of Cambridge, Madingley Road, Cambridge
CB3 0HA, UK
\and  Cavendish Laboratory, University of Cambridge, 19 J. J. Thomson Avenue, Cambridge
CB3 0HE, UK 
\and 
Dipartimento di Fisica e Astronomia, Università degli Studi di Bologna, Via Piero Gobetti 93/2, I-40129 Bologna, Italy
\and 
European Southern Observatory, Karl-Schwarzschild-Str. 2, D-85748 Garching b. München, Germany
\and
Observatoire de Genève, Université de Genève, 51 Ch. des Maillettes, 1290, Versoix, Switzerland 
\and
Laboratoire AIM-Paris-Saclay, CEA/DSM/Irfu, CNRS France 
\and
Department of Astronomy, The University of Texas at Austin, Austin, TX 78712, USA
\and 
INAF - Astronomical Observatory of Trieste, via G.B. Tiepolo 11, I-34143 Trieste, Italy
\and 
N'ucleo  de  Astronom'ia,  Facultad  de  Ingenier'ia,  Universidad  Diego
Portales,  Av.    Ej'ercito  441,  Santiago,  Chile
\and
Department of Physics and Astronomy, University College London, Gower Street, London WC1E 6BT, UK
\and
Astrophysics, The Denys Wilkinson Building, University
of Oxford, Keble Road, Oxford, OX1 3RH
\and
Kapteyn Astronomical Institute, University of Groningen, Postbus 800, 9700 AV, Groningen, The Netherlands
\and
Max  Planck Institut f\"ur Extraterrestrische Physik Giessenbachstrasse 1, Garching D-85748, Germany
\and
 Imperial college, Kensington, London SW7 2AZ, UK
\and 
European Southern Observatory, Avenida Alonso de Córdova 3107, Vitacura, 19001 Casilla, Santiago de Chile, Chile
\and
School of Physics and Astronomy, University of Nottingham, University Park, Nottingham NG7 2RD, UK 
\and
University Observatory Munich, Scheinerstrasse 1, D-81679 Munich, Germany
\and
Department of Astronomy, University of Michigan, 311 West Hall, 1085 South University Ave., Ann Arbor, MI 48109-1107, USA
\and
Instituto de Astrofísica e Ciências do Espaço, Universidade de Lisboa, OAL, Tapada da Ajuda, P-1349-018 Lisbon, Portugal
\and 
Departamento de F\'{i}sica, Faculdade de Ci\^{e}ncias, Universidade de
Lisboa, Edif\'{i}cio C8, Campo Grande, PT1749-016 Lisbon, Portugal
\and
Instituto de Fisica y Astronomia, Facultad de Ciencias, Universidad de Valparaiso, 1111 Gran Bretana, Valparaiso, Chile
\and
Institute d'Astrophysique de Paris, CNRS, Université Pierre et Marie Curie, 98 bis Boulevard Arago, 75014, Paris, France
\and
INAF - Osservatorio Astrofisico di Arcetri, Largo E. Fermi 5, I-50157 Firenze, Italy
\and
National Optical Astronomy Observatory, 950 North Cherry Ave, Tucson, AZ, 85719, USA
\and
UCO/Lick Observatory, Department of Astronomy and Astrophysics, University of California, Santa Cruz, CA 95064, USA
\and
Harvard-Smithsonian Center for Astrophysics, 60 Garden St, Cambridge MA 20138
\and
Space Telescope Science Institute, 3700 San Martin Drive, Baltimore, MD, 21218, USA
\and 
Dark Cosmology Centre, Niels Bohr Institute, University of Copenhagen, Juliane Maries Vej 30, DK-2100 Copenhagen, Denmark
\and
Astronomy Department, University of Massachusetts, Amherst, MA 01003, USA 
\and
 Pontificia Universidad Católica de Chile  Instituto de Astrofísica 
Avda. Vicuña Mackenna 4860 - Santiago - Chile 
 \and
Aix Marseille Universit\'e, CNRS, LAM (Laboratoire d'Astrophysique  de Marseille) UMR 7326, 13388, Marseille, France 
\and
Instituto de Astrofísica de Canarias, Calle Vía Láctea s/n, E-38205 La Laguna, Tenerife, Spain
\and
Departamento de Astrofísica, Universidad de La Laguna, E-38200 La
Laguna, Tenerife, Spain
\and
 Faculty of Physics, Ludwig-Maximilians Universität, Scheinerstr. 1, 81679, Munich, Germany
\and
 Excellence Cluster, Boltzmannstr. 2, D-85748 Garching, Germany
\and
Department of Physics, Durham University, South Road, DH1 3LE Durham, UK
\and 
Dipartimento di Fisica Universitá degli Studi di Ferrara, via Saragat 1, I-44122 Ferrara, Italy
\and
Leiden Observatory, Leiden University, 2300 RA, Leiden, The Netherlands 
\and
Steward Observatory, The University of Arizona, 933 N Cherry Ave, Tucson, AZ, 85721, USA
\and
Department of Physics and Astronomy, PAB, 430 Portola Plaza, Box 951547, Los Angeles, CA 90095-1547, USA
\and
School of Physics and Astronomy, University of St. Andrews, SUPA, North Haugh, KY16 9SS St. Andrews, UK}

\abstract{
This paper describes the observations and the first data release (DR1) of the ESO public spectroscopic survey "VANDELS, a deep VIMOS survey of the CANDELS CDFS and UDS fields". 

VANDELS'  main targets are  star-forming galaxies at redshift $2.4<z<5.5$, an epoch when the Universe was less than 20\% of its current age,  and massive passive galaxies in the range $1<z<2.5$. By adopting a strategy of ultra-long exposure times, ranging from a minimum of 20 hours to a maximum of 80 hours per source, VANDELS is specifically designed to be the deepest ever spectroscopic survey of the high-redshift Universe. Exploiting the 
red sensitivity of the refurbished VIMOS spectrograph, the survey is obtaining ultra-deep optical spectroscopy covering the wavelength range 4800-10000 \AA\ with sufficient signal-to-noise to investigate the astrophysics of high-redshift galaxy evolution via detailed absorption line studies of well defined samples of high-redshift galaxies.

The VANDELS-DR1 is the release of all medium-resolution spectroscopic data obtained during the first season of observations, on a 0.2 square degree area centered around the  CANDELS-CDFS  and CANDELS-UDS areas. It includes data for all galaxies for which the total (or half of the total) scheduled integration time was completed. 
The data release contains 879 individual objects, approximately half in each of the two fields, which have a measured redshift, with  the highest reliable redshifts reaching  $z_{spec} \sim 6$. 
In the data release we  include fully wavelength and flux-calibrated 1D spectra, the associated error spectrum and sky spectrum and  the associated wavelength-calibrated 2D spectra. We also provide  a catalogue with the essential galaxy parameters, including spectroscopic redshifts and redshift quality flags measured by the  collaboration. 
In this paper we present the survey layout and observations, the data reduction and redshift measurement procedure and the general properties of the VANDELS-DR1 sample. In particular  we discuss  the spectroscopic redshift distribution, the accuracy of the  photometric  redshifts for each individual target category and we provide some examples of data products for the various target types and the different quality flags. 

All VANDELS-DR1 data  are publicly available and can be retrieved from the ESO archive. Two further data releases are foreseen in the next two years with a  final data release currently scheduled  for June 2020 which will include 
improved re-reduction of the entire spectroscopic data set. 
}

\authorrunning{L. Pentericci et al.} 
\titlerunning{VANDELS}
    
  \date{Received ; accepted}
 \maketitle
 \section{Introduction}
 Making significant progress in our understanding of galaxy formation and evolution requires observations of substantial 
samples  of galaxies,  over  a  large  enough  volume  and a large range of mass and redshift. Only with representative samples and reliable observations we are able to test models of galaxy formation in a rigorous way.  In particular, spectroscopic surveys can play a key role,  since they provide, besides accurate information on the galaxies' redshift, a wealth of other important observable  properties, such as  emission and  absorption line features, measures of internal motions, spectral indices and spectral  breaks.   These in  turn allow  us to characterise the intrinsic physical properties of galaxies, the nature of their stellar populations including the chemical composition,  the presence of non thermal sources, the ionizing radiation field  and the  evolution of all these properties with cosmic time \citep{ellis17}.

In recent years a series of spectroscopic campaigns have been carried out, starting from the low redshift Universe where the Sloan Digital Sky Survey (SDSS) observed more than  a million galaxies at redshifts z $\sim$ 0.1-0.7 \citep{reid16,alam15}. At redshifts z $\sim$1 a series of spectroscopic surveys have sampled large volumes of the Universe,  by observing many tens of thousands  galaxies: in this redshift range, the Very Large  Telescope (VLT) and the VIsible MultiObject Spectrograph (VIMOS) have played  a major role with the the VIMOS Very Deep Survey (VVDS \citealt{lefevre05}), the COSMOS spectroscopic survey (zCOSMOS \citealt{lilly07})  and the VIMOS Public Extragalactic Redshift Survey (VIPERS  \citealt{guzzo14}). These surveys contributed substantially to improving our understanding of galaxy evolution also in relation to the environment.
At even higher redshift ($z> 2$)  spectroscopic surveys have been necessarily more limited with at most several thousand galaxies identified at z$\sim$ 2-4 (e.g. KBSS-MOSFIRE, \citealt{steidel14} and VUDS, \citealt{lefevre15}), which become less than few hundreds as we approach the reionization epoch at $z \geq 5$  (\citealt{debarros17,shy18}; Pentericci et al. 2018 subm).  The main aim of these surveys has been the redshift identification of increasingly distant (and faint) galaxies, particularly star forming objects. A somewhat complementary approach was employed by the Galaxy Mass Assembly ultra-deep Spectroscopic Survey (GMASS ,  \citealt{kurk13,cimatti08}), which   investigated the physical and evolutionary processes of galaxy mass assembly in the redshift range of $1.5<z<3$, by obtaining ultra deep optical spectra (up to 32 hours), which allowed a detailed spectral study of a small sample of  passive galaxies at $z\sim1.5$ and several tens of star forming galaxies up to $z\sim3$. 

VANDELS, a VIMOS survey of the CANDELS CDFS and UDS fields,  is an ESO public  spectroscopic survey   designed  to complement and extend the work of these previous campaigns by focusing on  ultra-long exposures  of a relatively small  number of galaxies, pre-selected to lie  at high redshift. VANDELS started observations in August 2015 and it was  completed in  February  2018.
Exploiting the red sensitivity  of the refurbished VIMOS spectrograph, and ultra-long integration times of up to 80 hours on source,  the survey is obtaining ultra-deep optical spectroscopy of around 2100 galaxies in the redshift interval $1.0 < z < 7.0$, including  star forming galaxies at redshift $z>2.4$,   an epoch  when the Universe was less than 20\% of its current age and massive passive galaxies in the range $1<z<2.5$. VANDELS has  observed  in the  wavelength range 4800-10000 \AA\ with intermediate resolution. 
VANDELS' prime motivation is to  move  beyond simple  redshift  acquisition, by obtaining  spectra  with  high  enough  signal-to-noise to  allow detailed absorption and emission  line studies in individual spectra,  derive  metallicities  and  velocity  offsets, and finally derive improved constraints on physical parameters, such as stellar  mass and star formation rates. These informations will    enable   a  detailed  investigation  of  the  physics  of  galaxies  in  the  early  Universe.   By targeting two extragalactic survey fields with superb multi-wavelength imaging data, including the best optical+nearIR+Spitzer imaging,   VANDELS 
will produce a unique legacy dataset for 
exploring the physics underpinning 
high-redshift galaxy evolution.

 In this paper, after a brief summary of the target selection (Section 2),  we present the survey layout and observations (Section 3), the data reduction process (Section 4), the redshift measurement procedure (Section 5) and, finally, the content of the  first data release (DR1) (Section 6)  with the general description of the data publicy available, the properties of the galaxies in the release and some examples of data products for  different types of galaxies. 
In a companion paper (McLure et al. 2018), we present complementary information including the main scientific motivations of the survey, the assembly of the photometric catalogs, the determination of the photometric redshifts  and the  selection for the different category  of targets  that were observed in the survey.

We refer to total magnitudes in the AB system \citep{oke}. 
When quoting absolute quantities we assume  a cosmology with $H_0 = 70$ km s$^{-1}$ Mpc$^{-1}$, $\Omega_\Lambda = 0.7$, and $\Omega_m = 0.3$. 
 \section{The VANDELS public spectroscopic survey}

We provide a  brief summary of the survey below, but  the reader is referred to  McLure et al. (2018) for a detailed description  of the target selection and survey definition. 

The VANDELS survey targets two fields, the UKIDSS Ultra Deep
Survey (UDS: 02:17:38, -05:11:55) and the Chandra Deep Field
South (CDFS: 03:32:30, -27:48:28). These fields were selected since their central areas  have the best available HST multi-wavelength data from the CANDELS treasury survey \citep{koek11,grogin11}, as well as a wealth of ancillary data including ultra-deep IRAC photometry. 

For  the CANDELS/HST  areas (CDFS-HST and UDS-HST)  we  adopted the
photometric catalogues, based on $H_{160}$-band detections,  provided by the CANDELS team \citep{galametz+13,guo+13} to select our targets. 
These catalogues provide PSF homogenised photometry for the available ACS and WFC3/IR imaging, in addition to spatial-resolution matched photometry from Spitzer
IRAC and key ground-based imaging data sets.

 Because of the large field of view of VIMOS, the spectroscopic observations cover also areas that are outside the original CANDELS footprints: for these  wider-field areas (CDFS-GROUND and UDS-GROUND) there were no near-IR selected, photometric catalogues therefore new multi-wavelength photometric
catalogues were assembled  using the publicly available imaging. The
assembly of the new catalogues and the multiband data employed,  is described in details in McLure et al. (2018).

For the two central regions covered by deep HST imaging,
 we adopted the photometric redshift solutions by the CANDELS survey team
\citep{santini+2015,guo+13}, which were  derived by optimally combining several  independent estimates produced by different photometric redshift codes by CANDELS team members, as described in details in \cite{dahlenetal2013}. For the wider areas outside the CANDELS footprint, new photometric redshifts were generated by our team, based on the new 
photometric catalogues described above. Fourteen  independent photometric redshifts estimates were generated by eleven individual team members, based on different codes and methods. They were then combined 
taking the median value of zphot for each galaxy and producing an official VANDELS zphot.

Target selection was then carried out using the CANDELS and the new VANDELS photometric catalogs and photometric redshifts, respectively for the areas with deep HST imaging and for the wider areas. Here we briefly recall the  main categories of objects that were selected:
\\
1) bright star forming galaxies (SFG): this sample consists of star forming galaxies with  $ i\leq 25$ with a best photometric redshift in the  range $ 2.4 \leq z_{phot} \leq 5.5$  so that the main absorption features necessary to investigate the metallicity (e.g. as in \citealt{somma12} and \citealt{rix04}) fall in the observed spectral range.  In practice the redshift range of the resulting  sample observed is limited to $z_{phot}<5$. The galaxies were required to have $sSFR>0.1 Gyr^{-1}$ where sSFR is the specific star formation rate derived by the  SED fitting described by  McLure et al. (2018), assuming the best photometric redshift.  
\\
2) passive galaxies (PASS): this sample consists of UVJ selected passive galaxies (see \citealt{williams09} and McLure et al. 2018 for more details on the definition) which have  a photometric redshift in the range  $1 \leq z_{phot} \leq 2.5$ with $H \leq 22.5$ and $i \leq 25$. The magnitude constraints is equivalent to a minimum total stellar mass of $10^{10} M_\odot$. In practise the redshift range of the resulting  sample observed is limited to $z\sim 2$. 
\\
3) fainter star forming galaxies that we call  Lyman Break Galaxies (LBG) to differentiate them from group 1: this sample consists of galaxies with a photometric redshift in the range  $3 \leq z_{phot} \leq 5.5$ which have $H \leq 27$ and $i \leq 27.5$ ( $i \leq 26$ in the wider regions) and galaxies with $5.5 \leq z_{phot} \leq 7$ which have $H \leq 27$ and $z \leq 26$ ($z \leq 26$ and   $z \leq 25$ in the wider regions, respectively for the UDS and ECDFS). The redshift range is such that the Ly$\alpha$ emission line or the Lyman break fall within the observed spectral range.  As for the bright star forming galaxies, the targets were required to have $sSFR>0.1 Gyr^{-1}$.

In the above cases the i-band (or z-band for objects with $z_{phot}>5.5$) constraints are imposed in order to ensure that the final 1d spectra have a minimum S/N in the observed spectral range around 6000-7000\AA, specifically S/N per resolution element $> 15$ for objects with $i<24.5$ and S/N per resolution element $\sim 10$ for the faintest $i=25$ objects. For the faint star-forming galaxies, the VANDELS strategy is designed to provide a consistent \lya\ emission line detection limit of $\sim 2 \times 10^{-18}$ erg s$^{-1}$ cm$^{-2}$ at 5$\sigma$. 
\\
In addition, a small sample  of Herschel detected sources with $z_{phot}>2.4$ and $i<27.5$ were selected in both fields (1\% of the sample), and in the CDFS we further added  another 2\% of targets selected as AGN, which were selected  thanks to the availability of the Chandra 4Msec \citep{xue} observations as in \cite{hsu}, or via
IRAC power-law and 24$\mu m$ detection as in \cite{chang}, with the further constraints of having 
 $z_{phot}>2.4$ and $i<27.5$.  For these, the
photometric redshifts present in our catalog where used. 

Following the above selection criteria a total of 9656, objects were available in the final catalogs, approximately half for the UDS and CDFS fields respectively. 
\section{Survey layout and observations}
Our survey is conducted with VIMOS \citep{lefevre2003} which is mounted on the ESO–VLT unit number 3 “Melipal" and  has a pixel scale of 0.205$''$/pixel with  a total field of view of 4 x 7$'$ x 8$'$. We used the  medium resolution  grism which  gives  a wavelength coverage from 4800 to 10000 \AA\ and provides  a  resolution of 580 with a dispersion of 2.5\AA/pix. Spectral multiplexing (i.e. the ability of having more than one slit along the dispersion direction) is possible only when sources are positioned at the very upper or bottom edge of the FOV. 
As detailed in McLure et al. (2018), the  VANDELS survey targets a total of eight VIMOS pointings, four pointings in the UDS field and four pointings 
in the CDFS field. Each VANDELS pointing has four associated masks, each of which is observed for 20-hours of on-source integration time. The survey utilizes a nested slit allocation policy, such that the 
brightest objects within a given pointing appear on a single 
mask (receiving 20-hours of integration), fainter objects appear on two masks (receiving 40-hours of integration) and the faintest objects appear
on all four masks (receiving 80-hours of integration). In this way we can reach a very similar S/N on the continuum (for the passive and star forming galaxies) or emission line flux limit (for the faint star forming galaxies)  for all our sources. In the next subsections we will describe the target allocation procedure and the procedure to make the VIMOS masks.

\subsection{Slit allocation}

Because of the  nested strategy that VANDELS employs, all target allocation was done simultaneously for  all masks at the beginning of the survey, in order to maximise the total number of sources observed.
Extensive simulations were run using the Slit Positioning  Optimization Code (SPOC, \citealt{bottini05}). 
The primary goal of the simulation was to maximize the total number of total  slits allocated to VANDELS targets, keeping particular attention to the two categories of  bright star-forming galaxies and massive passive galaxies, which are the samples  with the lowest surface densities. The only other  constraint applied during the simulation work was the requirement 
to allocate the slits to objects requiring 20, 40 and 80 hours of integration in approximately a 1:2:1 ratio. 
We did not apply any  additional prioritization (e.g. in terms of redshift or source brightness) during  the slit allocation.
Before the simulations it was decided to keep a minimum slit length of 7$"$ (28 pixels)  to facilitate sky background subtraction and that the targets should  be positioned at least 8 pixels from one edge, to take into account the nodding strategy that is employed in the observations. Targets were treated as point sources.

Given the uneven distribution in ra-dec of our targets, due also to the fact that only  the central area is covered by the deep CANDELS data (representing about 45\% of the total area),  the original goal stated in the proposal to observe  a total of  1280 objects per field  proved to be  too ambitious, and a more realistic set of numbers was about 20\% lower for all categories.  
We  notice that the total  area covered by the VIMOS 
pointings in the CDFS is ~ 20\% smaller than in the UDS, 360 sq arcmins instead of 460 sq arcmins in UDS,  due to a different choice of pointings centers.
The CDFS also has a surface density of PASS targets which is lower by $\sim$15\% compared to the UDS, due to cosmic variance. 
Finally, in  the CDFS  field we have the further set of AGN  selected targets  which is absent in the UDS, but this group  comprises only 2\% of the catalog.

In a first pass  we only took the  SFG+PASS+HERSCHEL targets as input (adding the AGN for the CDFS field) 
to check what is the maximum  number of such targets we could  observe. Then a series of runs were done using  this sample as a forced set of  targets and  adding  the list  of LBG galaxies, trying to maximize the total number of targets  observed, while avoiding to penalize too much the PASS+SFG targets.
The best solution was obtained with a total of 1078 galaxies for the UDS field (1028 for CDFS),  i.e. 693, 224, 151 and 10 respectively for LBG, SFG , PASS and Herschel for the UDS field  (656, 200,  117, 9 and 46 for  LBG, SFG, PASS, HERSCHEL and AGN for the CDFS field). 
A final further optimization was run using this set of  1079(1025) targets as input catalog to position slits but this time  allowing completed targets to still be in the observable pool of objects for subsequent pointings: in this way several tens of targets get a higher S/N than intially allocated although we lose the completion of 1 target for UDS (gain 3 targets for CDFS).
We finally remark that  although the HST regions represent about 45\% of the total survey area, the higher surface density of faint $z>3$ galaxies allowed us to maximize the number of slits in these central HST areas.
The HST selected galaxies are approximately 54\% and 48\% of the total targets respectively for the CDFS and UDS fields. 

\begin{figure*}
\includegraphics[trim={0 0 12cm 0},width=15cm,clip=]{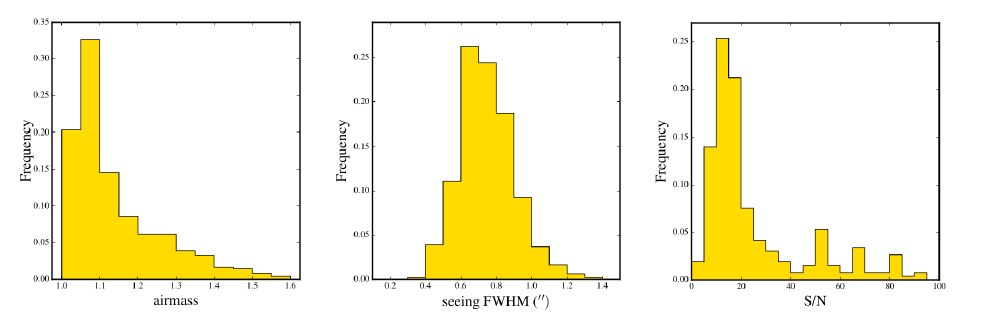}
\caption{The observing conditions under which the data released in DR1 were obtained: left is the airmass of individual exposures, mostly constrained below 1.5; right  is the seeing, measured directly on the spectra, mostly constrained below 1$''$}
\label{fig:stat}
\end{figure*}
\begin{figure}
\includegraphics[width=8cm]{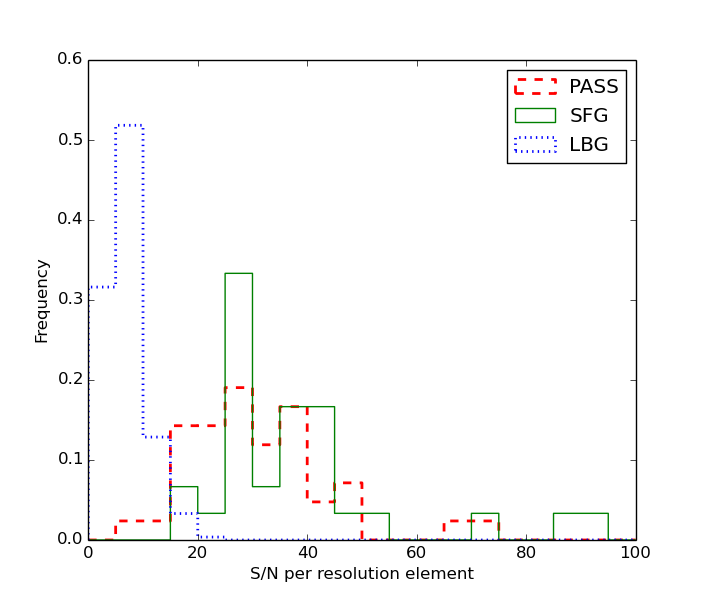}
\caption{The distribution of the median S/N per resolution element,  of all  completed DR1 spectra, measured in the range 6000-7400 \AA\ and separated for the three main VANDELS classes. The red dashed  histogram  refers to PASS galaxies, the blue dotted histogram to LBGs and the green histogram to SFGs. Each histogram is normalised separately.}
\label{fig:sn}
\end{figure} 
\subsection{Mask preparation}\label{masks}

Mask preparation was done using the VIMOS Mask Preparation Software
(VMMPS, \citealp{bottini05}), distributed by ESO. VMMPS requires the
acquisition of a direct image (pre-imaging), from which a list of
visible sources is extracted. The pre-imaging was obtained in October  2014 in service mode, in the R band. The catalogue is then cross-matched with the
catalogue of sources contributed by the user, so to match the
astrometry of the targets catalogue to the instrument coordinate
system. After this, the cross-matched position of each
VANDELS target was overplotted on the pre-image and visually
inspected. If there was a mis-match in RA or Dec of one pixel or more,
the target position was modified. This check has been done directly on
the targets which were bright enough to be visible in the preimaging data,
while for the faint targets the position of close-by brighter sources
was checked with respect to their pre-image counterparts, and the
possible mismatch used to modify the target position.  The same
procedure was applied to the two reference stars, chosen in each VIMOS
quadrant.

As a second step, VMMPS assigns the slits to the input targets. 
We imposed a slit length of at least 28 pixels which means that the target center must always be at least 14 pixels away from both slit ends, while its maximum length is normally maximised by VMMPS during the optimisation process, to allow for a better sky
subtraction. Given the  simulations described in Sect. 3.1,
for each mask, we had an input catalogue of targets which were
supposed to perfectly match the VIMOS multiplexing, given the slit
constraints and the chosen set up. For this reason, it was
expected that VMMPS could place  a slit on all the input targets. This was not
always the case for the following reasons: 
 i) slightly different
field of view with respect to the simulation, which in some cases made
us miss the target closest to the left border of the CCD; ii) the
initial astrometry match, which in crowded regions can modify the
target position enough not to match anymore the constraints on slit
length; iii) the presence of the two reference stars in each  quadrant (not considered during the simulations), which reduced the
area on the CCD available to slits.
For all the targets which were not automatically assigned a slit by
VMMPS, we verified if they could be fit in a slit by loosening the
constraints on the position of the target within the slit, still 
keeping a minimum distance of the target from the slit borders of at
least 10 pixels. This way we were able to fit in a slit almost all the
input targets. On average, we lost 1-2 targets per quadrant, mainly
because of the need for the  reference stars.

\subsection{Observations}

All VANDELS observations were obtained in visitor mode during a period which ran from August 2015 to February  2018. Spectra belonging to the Data Release 1 were observed during the first  observational season which ran  from August  2015 to February 2016.  The individual OBs were designed to deliver a total of one-hour of on-source 
integration time. Each OB consisted of three integrations of 1200s, obtained in a three-point dither pattern, with dither offsets of 0.82 
arcseconds (dither positions 0,-0.82$"$,+0.82$"$) corresponding to 4 pixels. This was done to  remove most of the small-scale detector pattern and facilitate sky-subtraction.
For the same reason we have tried to obtain an equal number of frames at each of the three positions, even if on individual nights it was not always possible to complete the  OBs.

One arc and one flat exposures were obtained for calibration after the execution of OB: it was  possible to perform one calibration every two OBs and save a little bit of time skipping the second calibration. In practise  the observing sequence was set to be: science-calib-science.
This was true for all the observations done while the target was  rising/setting. If in one OB the target was  rising and in the following one the target was  setting we have obtained calibration at the end of both OBs.   Finally, a spectrophotometric standard star was observed at least once every 7 nights and at least  once per run during twilight, if possible under photometric conditions. Because of our wide wavelength coverage, we used  bright late-type  (F-G) stars such as 
LTT9239 and LTT3864 that cover up to 1$\mu m$.

The nominal observing conditions required for a single exposure to be validated were: moon illumination $\leq 0.5$, moon angular distance above 90$^{\circ}$, seeing $\leq 1.0 "$ FWHM, as measured directly on the spectra of the brightest objects which were visible in the single exposures, airmass $\leq 1.5$ and clear weather  conditions. An exposure was still validated if one (and only one) of the above conditions was not met,  but the discrepancy was   less than 20\% (e.g. seeing $\leq 1.2''$), while  all other conditions were satisfied. The visiting  observers judged directly from the quality of the spectra if the weather could be considered clear/photometric/thin (in some cases overriding the official ESO conditions) from the S/N of the brightest objects in the masks which were visible in 20 minute exposures.
 
In Fig. \ref{fig:stat}  we show  histograms of 
airmass and  FWHM  for all individual exposures  obtained during season one. The median seeing of the observations is 0.7$"$ and only a
tiny fraction of the data is obtained with seeing $\geq 1''$; similarily a negligible fraction of the data has been obtained at airmass >1.5. In Figure \ref{fig:sn} we show the distribution of signal-to-noise ratio (S/N) for the completed spectra released in DR1. The S/N  is  determined from the error spectrum  as the median value  in the range 6000-7400 \AA, which is essentially free from bright skylines,  per spectral resolution element (one spectral resolution element is equal to $\sim$4 pixels). The distribution of S/N obtained agrees very well 
with the S/N predictions we made in the original proposal.
\begin{figure*}
\includegraphics[width = 16cm,clip=]{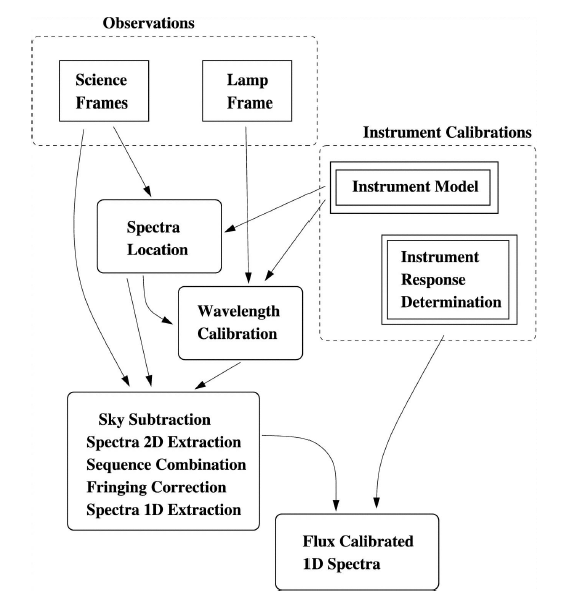}
\caption{A flow diagram illustrating the various steps of the data reduction pipeline }
\label{fig:red}
\end{figure*}

\section{Data reduction}
The VANDELS  data was reduced  using  the  fully-automated pipeline Easy-Life which starts from  the raw data and produces wavelength-and flux-
calibrated spectra. The pipeline 
is an updated version of the algorithms and dataflow from the original VIPGI system, fully  described in \cite{scodeggio05}. 
In Figure \ref{fig:red} we show a flow diagram illustrating the key features of the data reduction pipeline.
The first step in the reduction of VIMOS science data is the canonical preliminary reduction of the CCD frames, which includes pre-scan level and average bias frame subtraction, trimming of the frame to eliminate pre-scan and over-scan areas, interpolation to remove bad CCD pixels and flat  fielding. After the preliminary reduction step, subsequent data reduction steps are carried out on all MOS slits individually. For each individual spectroscopic exposure the wavelength calibration derived from the arc exposures is checked against the positions of bright sky lines and the local 
inverse dispersion solution modified to account for any discrepancies.
The next steps in the data reduction procedure are the object detection and sky subtraction for each MOS slit spectrum within each individual 1200s science exposure. Initially, the slit spectrum is collapsed along the wavelength axis, following the geometrical shape defined by the local curvature model, to produce a slit cross-dispersion profile.  A robust determination of the average  signal level and r.m.s in this profile is then obtained using an iterative 
$\sigma$-clipping procedure, and 
objects are detected as groups of contiguous pixels above a given detection threshold. Before wavelength calibration is applied, a median estimate of the sky spectrum, derived using all the 
pixels that are devoid of object signal, is subtracted from each slit. 
The sky spectrum is estimated separately for each individual science exposure due to the significant variation in OH line 
strength over the timescale of a typical spectroscopic exposure. 
The sky-
subtracted slit spectra are then two-dimensionally extracted
using the tracing provided by the slit curvature model, and resampled to a common linear wavelength scale. Only after this point are the single exposures of a pointing combined together.  First, the N two-dimensionally extracted spectra for each slit are median combined (with object pixels masked), without taking into account the jitter off-sets, to produce a two-dimensional sky-subtraction residual map. 
The residual map is then subtracted from all the N two-dimensional single-exposure slit spectra,  improving the sky-subtraction and removing any residual fringing. At this point a second combination is carried out, this time taking into account the jitter offsets among the N individual two-dimensional slit spectra (as determined during the previous object detection procedure).

The single-exposure, residual-map subtracted, spectra are off-set to compensate for the effect of the jitter, and a final average two-dimensional spectrum for each slit is obtained. The object detection process is repeated on the combined two-dimensional spectra to produce  the final catalogue of detected spectra, and a one-dimensional spectrum is extracted for each detected object, using the Horne optimal  extraction procedure \cite{horne86}. Finally, spectra are  flux calibrated using a simple polynomial fit to the instrument response curve derived from observations of spectrophotometric standard stars, and corrected for telluric absorption features. 
The last  correction is based on a template absorption spectrum derived for each combined jitter  sequence from the data themselves. 
The  final flux calibration was  performed correcting the spectra  for both atmospheric and galactic extinction and then normalizing them  to the i-band photometry available for each target. This procedure 
was already successfully  employed by the VIMOS Ultra Deep Survey \citep{lefevre15}. We plan to improve the calibration procedure employing additional broad band filters: the final data release will feature a re-reduction of the entire spectroscopic data set, incorporating this and possibly other improvements. 
\subsection{Problems with blue-end calibration}
\begin{figure}
\includegraphics[width = 10cm,clip=]{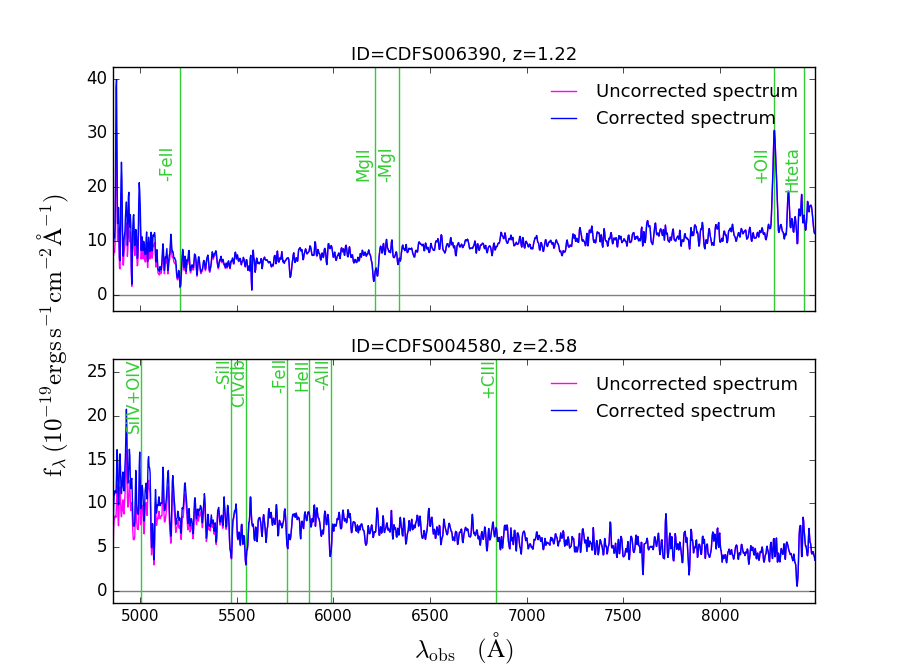}
\caption{The effect of the blue flux correction: in blue the final corrected spectrum and in magenta the original flux calibrated spectrum  for CDFS006390 at z=1.22 (top panel)  which was selected in the PASS sample, and CDFS004580 at z=2.58 which was selected in the SFG sample. We also indicate with green vertical lines the main spectral features identified. }
\label{fig:corr}
\end{figure}

During    final    testing    of    the    flux    calibration    of    the    DR1    spectra,    it    became    clear    that    the    extreme    
blue    end    of    the    spectra    (i.e.  $\lambda \leq 5600 \AA$)
suffer    from    a    systematic    drop    in    flux    when    compared    to the    
available broad band  photometry.    The    underlying    cause    for    this    loss    of    blue    flux    is    still    under    investigation.
For    the    purposes    of the first data    release    we    have    implemented    an    empirically    derived    correction    to    
the    spectra    at  these blue wavelengths
which   accounts for   the    flux    loss
on    average. The empirical    correction,    which    has    been    applied    to    all    of    the    DR1    spectra, is    designed    to    ensure    that    the    final    spectra    of    bright    star‐forming    galaxies    in    the    redshift    interval    $2.4<z<3.0$    display    the    
expected    power-law    continuum    slopes    in    the    rest-frame    wavelength    range    ($1300\AA < \lambda < 2400 \AA$),
which    are    independently    confirmed    from    the    available    photometry. At    the    time    of    the    data    release,    we    believe    that    the    spectra    including    the    correction    for    blue    flux   
loss    represent    our    best    calibration    of
the    VANDELS    spectra.    However,    for    
completeness,    we    are    also    making    available    the    spectra    without    the    blue    flux    correction.
To show the effect of this blue correction, in Figure \ref{fig:corr} we present  two examples of spectra belonging to the SFG and PASS samples, respectively, and we compare the original and corrected flux. The effect is only noticable in the first few hundreds \AA\ of the spectral range, and it is most important for the star forming galaxies.

\section{Redshift measurement}
Spectroscopic redshifts and associated quality flags have been determined for all objects using the Pandora software 
package, within the EZ environment \citep{garilli10}. This software was used  sucessfully in previous VIMOS survey and can simultaneously  display  the 1D extracted spectrum, 
the 2D linearly resampled spectrum, the 1D sky spectrum and the noise. It is also possible to inspect the image thumbnail of the object with the exact position of the slit, to check for the presence of other 
sources in the same slit and the pixels over which the source was extracted.
The core algorithm for redshift determination is the correlation with available galaxy spectral templates.
A key element for the cross-correlation engine to deliver a robust measurement is the availability  of reference templates that cover a wide range of galaxy and star types and a wide range of rest-frame
wavelengths. To determine the VANDELS spectroscopic redshifts we adopted templates derived from previous VIMOS observations for the 
VVDS \citep{lefevre13} and zCOSMOS surveys \citep{lilly07}, with and without Ly$\alpha$ emission.  Alternatively, it is also possible to determine the redshift by estimating manually the center of one or more emission and aborption lines. 
In several cases, it was necessary to manually perform some cleaning of the spectra, i.e. removing obvious noise residuals at the location of strong sky lines, or the zero-order projection.

Each target was  assigned to two measurers
from the  VANDELS team  who independently determined the redshift and located the main spectral features (in emission or absorption).  
Each  measurer  also assigned a spectroscopic quality flag to the target:  these quality flags have been  allocated according to the original system devised for the VVDS and are related to the confidence 
of the spectral measurement. The reliability flag may take the following values:
\\
0: No redshift could be assigned (redshift is set to Nan)
\\
1: 50\% probability to be correct. Some low S/N lines are identified  but there is a weak to moderate match with templates.
\\
2: 70-80\% probability to be correct. There are several matching absoption lines and a general good match  with templates.  
\\
3: 95-100\% probability to be correct. The spectrum shows multiple  strong absorption and/or emission lines giving a consistent redshift and has a moderate S/N.  There is a strong cross-correlation signal with an excellent continuum match to templates. 
\\
4:100\% probability to be correct. The spectrum shows  multiple  strong absorption and/or emission lines giving a consistent redshift, and has  a high S/N. There is a strong cross-correlation signal with an excellent continuum match to templates 
\\
9: spectrum with a single emission line. The redshift given is the most probable given the  observed continuum and the shape of the emission line, and it has a >80\% probability to be correct. 
\\
We emphasize that the quality flag only reflects the accuracy of the redshift determination and in principle is not related to the S/N of the spectrum although almost all the QF=4 spectra have very high S/N. 
The quality flags for AGN spectra are preceded by a leading  1 (e.g 12, 14 etc), the quality flags for spectra which  were not primary targets (i.e. were serendipitously observed in  a certain slit)  are preceded by an additional 2; finally  the quality flags  for spectra deemed to be problematic are preceded by an additional  -1. In the DR1 only the AGN flags are present, since the secondary targets will be released only at the end of the survey.
Following their independent redshift determinations, the two measurers were required to compare  their redshifts and flags and to reconcile any differences.   As a final step, all spectra were 
also independently re-checked by the two PIs and any remaining discrepancies in the redshifts and quality  flags were again reconciled.
This final pass was especially necessary to homogenize the quality 
flags as much as possible,  given the different expertise and ability of the various redshift measurers. Based on repeated measurements, the typical accuracy of the spectroscopic redshift measurements is estimated to be +/-0.0005 ($\sim 150$ km s$^{-1}$).

There are 42 objects in the Data Release 1  with a previously published spectroscopic redshift. These were included in the survey since the expected integration time (mostly 80 or 40 hours) would result in spectra with a much higher S/N compared to those already available. Of these 42 sources,  one has no redshift assigned in VANDELS (but it is a source with half of the final scheduled integration) and one (AGN type) has a discrepant redshift, 3.442 with flag 1 in VANDELS and 2.448 from  \citep{trump13}, also with a low quality flag indicating that the redshift was based on the match with the photometric redshift.
For the other 40 objects, including all  quality flags both for VANDELS redshifts and for  the previous redshifts,  $\Delta z= (z_{VANDELS}- z_{old})/(1+z_{VANDELS})$  has a mean  value of -0.0003 and an rms of 0.0016, slightly higher than the redshift uncertainty quoted above.

\begin{figure}
\includegraphics[width = 9cm,clip=]{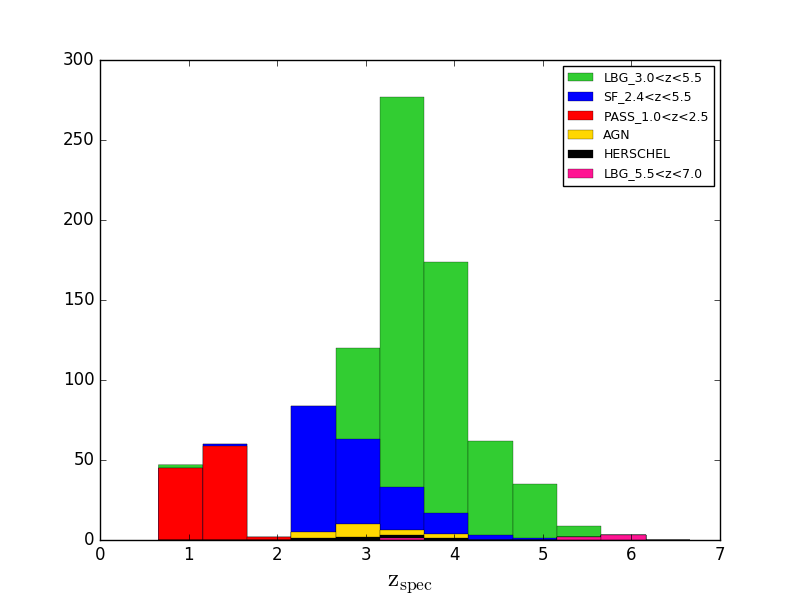}
\caption{The spectroscopic redshift distribution of all DR1 targets, divided by the original target classification}
\label{zspecdis}
\end{figure}

\begin{table*}
\begin{center}
\caption{Detailed list of entries in the catalogues associated to the data release 1 }
\begin{tabular}{llccccc}
\hline
Name & Description & Data type & Unit \\
\hline
ID   & The object ID & Char & \\
alpha & RA (J2000)  & Double & deg  \\
delta & Dec (J2000) & Double & deg  \\
$i_{AB}$ & The i-band magnitude  & float  & mag \\
$i-FILTER$ & The i-band filter   & string & \\
$z_{AB}$ & The z-band magnitude  & float  & mag \\
$z-FILTER$ & The z-band filter   & string & \\
$H_{AB}$ & The H-band magnitude  & float  & mag \\
$H-FILTER$ & The H-band filter   & string & \\
$t_{schedtime}$ & The  scheduled integration time & int & s \\
$z_{phot}$ & The  photometric redshift & float \\
$z_{spec}$ & Spectroscopic redshift & float \\
$z_{flg}$  & The quality flag for the redshift   & float \\
FILENAME & The fits filename of the spectrum & string \\
$t_{exptime}$   & The current integration time  & float  & s \\
\hline\hline
\end{tabular}
\label{tab:list}
\end{center}
\end{table*}
\begin{sidewaystable*}
\small{
\begin{center}
\tabcolsep=0.09cm
\scalebox{0.94}{
\begin{tabular}{lllllllllllllllcccccccccccc}
\hline 
ID & alpha &delta &$i_{AB}$ & i\textunderscore FILTER & $z_{AB}$ & z\textunderscore FILTER & $H_{AB}$ & H\textunderscore FILTER & $t_{schedtime}$ &  $z_{phot}$ & $z_{spec}$ & $z_{flg}$  & FILENAME & $t_{exptime}$  \\
\hline
 VANDELS\textunderscore UDS\textunderscore 000090 & 34.4182011 & -5.2764185 & 25.41889 & SUBARU\textunderscore i' & 25.59937 & SUBARU\textunderscore z' & 24.97066 & HST\textunderscore F160W & 144000  &    3.35 & 3.239 & 3.0 & VANDELS\textunderscore UDS\textunderscore 000090.fits & 71999.88 \\

  VANDELS\textunderscore UDS\textunderscore 000129 & 34.4883644 &-5.2760994& 25.73422&  SUBARU\textunderscore i'& 25.85772&  SUBARU\textunderscore z' & 25.77788 &  HST\textunderscore F160W & 144000 &      3.44 & 3.639 & 4.0 & VANDELS\textunderscore UDS\textunderscore 000129.fits & 72599.72 \\
  
 VANDELS\textunderscore UDS\textunderscore 021074 & 34.383926 & -5.1597713 & 25.45468 & SUBARU\textunderscore i'& 25.53815 & SUBARU\textunderscore z'&24.77221& HST\textunderscore F160W & 144000  &    3.01 &3.3044&  1.0 & VANDELS\textunderscore UDS\textunderscore 021074.fits & 72600.95 \\
  VANDELS\textunderscore UDS\textunderscore 389704&  34.4310368 &  -5.08566940 & 24.65722 & SUBARU\textunderscore i' & 24.57985 &  SUBARU\textunderscore z' & 24.17536 &  WFCAM\textunderscore H &  288000  &     3.58 &  3.5899&  4.0 &  VANDELS\textunderscore UDS\textunderscore 389704.fits & 143999.7 \\

  VANDELS\textunderscore CDFS\textunderscore 000541 &53.1066882 &-27.93019& 26.26715 & HST\textunderscore F775W &24.67075  &HST\textunderscore F850LP & 24.71833 &  HST\textunderscore F160W & 72000   &    5.764 &5.784 & 4.0 & VANDELS\textunderscore CDFS\textunderscore 000541.fits &78001.62 \\

   VANDELS\textunderscore CDFS\textunderscore 021735&  53.0628228 & -27.7264618 & 23.77141 &  HST\textunderscore F775W  &   23.00993  & HST\textunderscore F850LP & 21.08368 &  HST\textunderscore F160W & 144000    &   1.526 & 1.6087 & 2.0 &  VANDELS\textunderscore CDFS\textunderscore 021735.fits & 72001.39  \\

VANDELS\textunderscore CDFS\textunderscore 246958 & 53.2954379 &-27.681040 &        25.44884&  SUBARU\textunderscore IA738 & 25.17301 &HST\textunderscore F850LP & 26.77869& VISTA\textunderscore H&  144000 &     3.2 &    3.428 
& 9.0 & VANDELS\textunderscore CDFS\textunderscore 246958.fits & 149999.6 \\

  VANDELS\textunderscore CDFS\textunderscore 129520&  53.2628908&  -27.7349146 & 24.60112 &  SUBARU\textunderscore IA738 & 22.93470 &  HST\textunderscore F850LP & 20.77475&  VISTA\textunderscore H &   288000  &     1.59 &  1.6074&  3.0 &  VANDELS\textunderscore CDFS\textunderscore 129520.fits & 149999.6 \\
  ..... \\ 
\hline\hline
\end{tabular}}
\vskip0.2cm
\caption{Data release 1 catalogue: random examples of galaxies in the catalogue of the first data release. the entries are explained in Table 1. The data release is available at  http://www.eso.org/sci/publications/announcements/sciann17068.html and the catalogue is available at  https://www.eso.org/qi/catalogQuery/index/212  }
\label{tab:example}
\end{center}}
\end{sidewaystable*}

\section{VANDELS DR1-data}
The data release 1 (DR1) is available in the ESO archive and consists of all spectra obtained during the first VANDELS observing season, which  ran from August 2015 until February 2016. The data were acquired during runs 194.A-2003(E-K). 
The data release includes 
the spectra for all galaxies for which the final scheduled integration time was completed during season one (356 objects including 6 that actually received more than the scheduled time).
In addition, the data release also includes the spectra for 523 galaxies for which the scheduled integration time was 50\% complete at the end of season one (i.e. 20 out  of 40 scheduled  hours and  40 out of 80 scheduled hours). 
The total number of spectra released is  879 (415 in CDFS and 464 in UDS).
\\
For each target the following data files are being released:
\\
$\bf 1)$ the one-dimensional spectrum in FITS format, containing the 
following arrays
\\
WAVE: wavelength in Angstroms (in air)\\
FLUX: 1D spectrum blue-corrected flux in erg cm$^{-2}$ s$^{-1}$ angstrom$^{-1}$\\
ERR: noise estimate in erg cm$^{-2}$ s$^{-1}$ angstrom$^{-1}$\\
UNCORR-FLUX: 1D spectrum flux uncorrected for blue flux loss (see details in subsection 4.1)\\
SKY: the subtracted sky in counts
\\
$\bf 2)$  the two- dimensional resampled and sky subtracted (but not flux calibrated and atmospheric extinction corrected) spectrum in FITS format.
\\
$\bf 3)$ a catalogue with essential galaxy parameters, listed in Table \ref{tab:list}  and presented in Table \ref{tab:example} which include $i_{AB}$, $z_{AB}$ and $H_{AB}$ magnitudes with the relevant filters (see below), the scheduled and current integration times, and the spectroscopic redshift and quality flag determined as in Section 5.
As already mentioned  in Section 2, the UDS and CDFS fields are covered by different sets of observations, with 45\% of the area covered by deep  HST imaging (the CANDELS footprint) and the rest covered only by ground based imaging (the wider surrounding areas).  For this reason  the  $i_{AB}$, $z_{AB}$ and $H_{AB}$   magnitudes listed in the release catalogues are  generated from different filters. For each object, the origin of the $i_{AB}$, $z_{AB}$ and $H_{AB}$ photometry is listed in the catalogue in the columns with the filter name.
Here we list the match between  the catalogue photometry and  the filters used: \\
$\bullet$: $i_{AB}$ magnitudes refer to the SUBARU  i’-filter for the UDS-GROUND and UDS-HST targets, to the F775W filter for CDFS-HST targets and to the SUBARU IA738 filter for the CDFS-GROUND targets.\\
$\bullet$ $z_{AB}$ magnitudes refer to the SUBARU  z’-filter for the UDS-GROUND and UDS-HST targets and to the F850LP 
filter for the CDFS-GROUND and CDFS-HST targets \\
$\bullet$
$H_{AB}$ magnitudes refer to the F160W filter for the UDS-HST and CDFS-HST targets, the WFCAM H-filter for the UDS-GROUND targets and the 
VISTA  H-filter for the 
CDFS-GROUND targets

In the catalogue we report both the  total 
requested integration time ($t_{schedtime}$) and the current total integration time  ($t_{exptime}$). Clearly the objects for which the observations are not completed will have $t_{exptime} < t_{schedtime}$.
In some cases objects have $t_{exptime} > t_{schedtime}$: this is because,
due to changing observing conditions, some masks in both the UDS and CDFS fields have received slightly  more than their nominally scheduled 20 hours of on-source integration. In addition,   to optimize slit allocation,  six objects in this  data release 
requiring 20 hours of on-source integration
were placed on two VIMOS masks, and 
therefore actually received 40 hours of on-source integration.

\section{Galaxies in the Data Release 1} 
 \subsection{General properties and redshift distribution}
The targets in the DR1 belong to the following category: 
177 objects are  selected  as SFG with $2.4<z_{phot}<5.5$, 106 objects are PASS with $1.0<z_{phot}<2.5$, 566 objects are LBG with $3.0<z_{phot}<5.5$ and 6 are LBG with $5.5<z_{phot}<7.0$. In addition we have 18 targets selected as radio or X-ray AGN (all in the CDFS field) and 6 Herschel selected sources (3 in each field). These numbers, with the subdivision into fields are reported in Table \ref{tab:type}.

\begin{table}
\begin{center}
\caption{Galaxies in DR1 divided by target type}
\begin{tabular}{lccc}
\hline
Sample    & $N_{UDS}$ & $N_{CDFS}$ & $Ntot$   \\
\hline 
SFG 2.4<$z_{phot}$<5.5   &   110 & 67  &  177  \\
LBG 3.0<$z_{phot}$<5.5  &   285 & 281 &  566  \\   
PASS 1.0<$z_{phot}$<2.5 &   64  & 42  &  106  \\
LBG 5.5<$z_{phot}$<7.0  &    2  &  4  &   6 \\      
AGN            &    0  & 18  &  18 \\ 
HERSCHEL       &    3  &  3  &   6 \\
\hline
TOTAL          &   464 & 415 & 879 \\
\hline\hline
\end{tabular}
\label{tab:type}
\end{center}
\end{table}

\begin{table}
\begin{center}
\caption{Galaxies in DR1 divided by flag}
\begin{tabular}{lcccccc}
\hline
Flag     & SFG & LBG  & PASS & AGN & HERSCH & Total    \\
\hline 
Flag 0   &   0  &   6 & 0   & 0   & 0 & 6 \\
Flag 1   &  10  & 152 & 3   & 10  & 5 & 180   \\
Flag 2   &  19  & 127 & 5   & 5   & 0 & 156  \\   
Flag 3   &  29  & 137 & 18  & 0   & 0 & 184  \\
Flag 4(14)&  119 & 99  & 80  & 1   & 0 & 299   \\      
Flag 9   &  0   & 51  & 0   & 2   & 1 & 54    \\ 
\hline
  TOTAL       &  177 & 572 & 106 & 18  & 6 & 879 \\
\hline\hline
\end{tabular}
\label{tab:flag}
\end{center}
\end{table}

In Figure \ref{zspecdis} we present the redshift distribution of all DR1 spectra,  separated  by the original target classification (i.e. the AGN class includes the targets  originally selected as AGN and not the flag 14 objects).  
The quality flag statistics for galaxies in DR1 are reported in Table \ref{tab:flag} and are as follows: 299 galaxies i.e. 34\% of the sample,  received a flag 4 or 14; 184 i.e. 21\%, received a flag 3: 156 i.e., 18\%, received a flag 2; 180 i.e. 20\% received a flag 1; and 54 i.e. 6\%, received a flag 9. For 6 objects no redshift was assigned and therefore they are labelled as flag 0. 
From the same Table we also see that for  the two main classes of targets (SFG and PASS) the large majority of galaxies has flag $ \geq 3$, while AGN and Herschel selected sources have all lower quality flags.
4 targets have received a flag 14 indicating that their spectra show AGN features: of these only one was originally selected as AGN, while two were selected as LBGs and one as SFG. In the other selected AGN, no prominent AGN features were identified even when the redshift was determined.

\subsection{Photometric redshift accuracy}
\begin{figure*}
\includegraphics[width = 10cm,clip=]{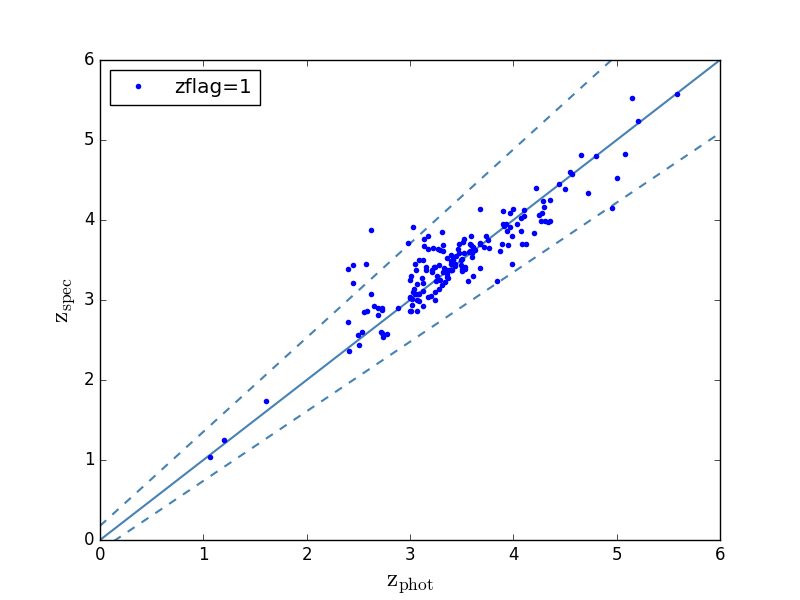}
\includegraphics[width = 10cm,clip=]{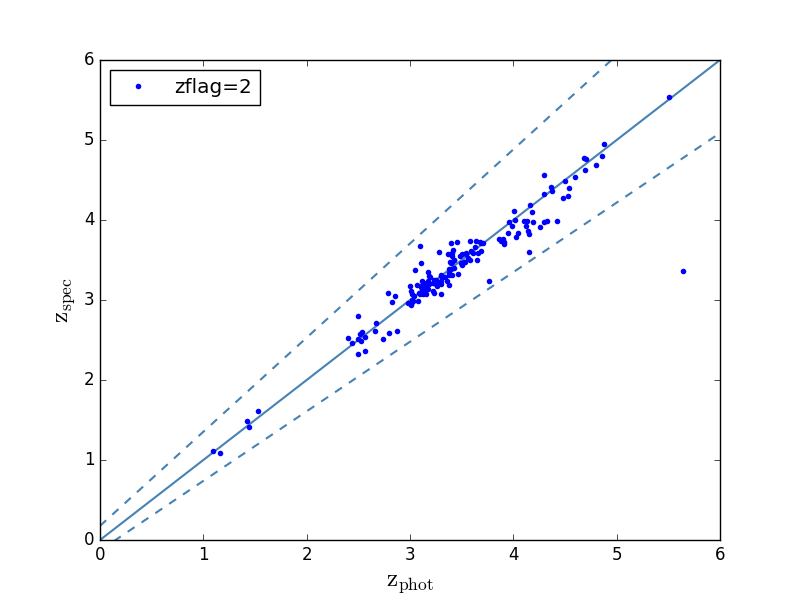}
\includegraphics[width = 10cm,clip=]{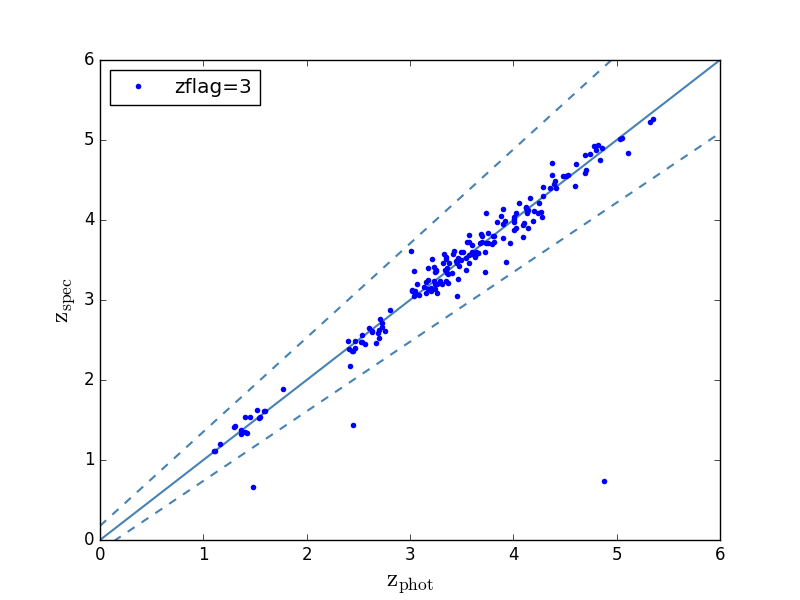}
\includegraphics[width = 10cm,clip=]{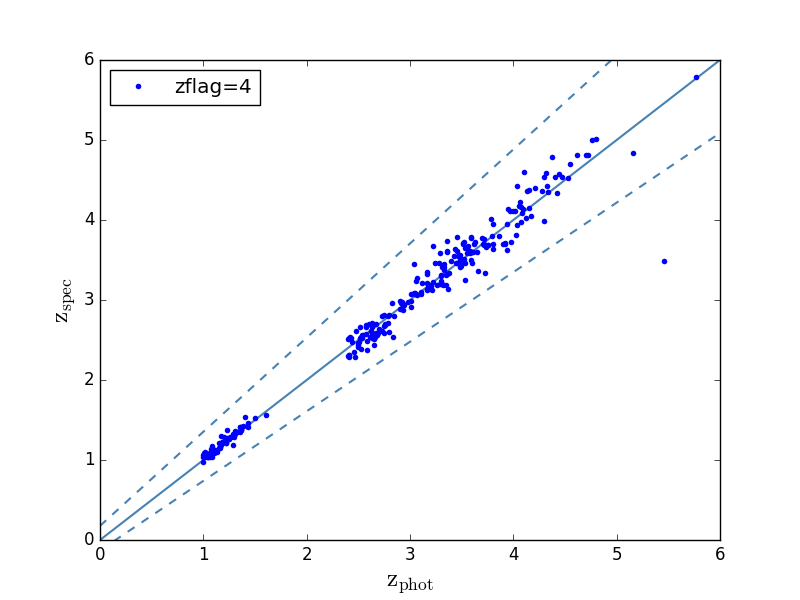}
\includegraphics[width = 10cm,clip=]{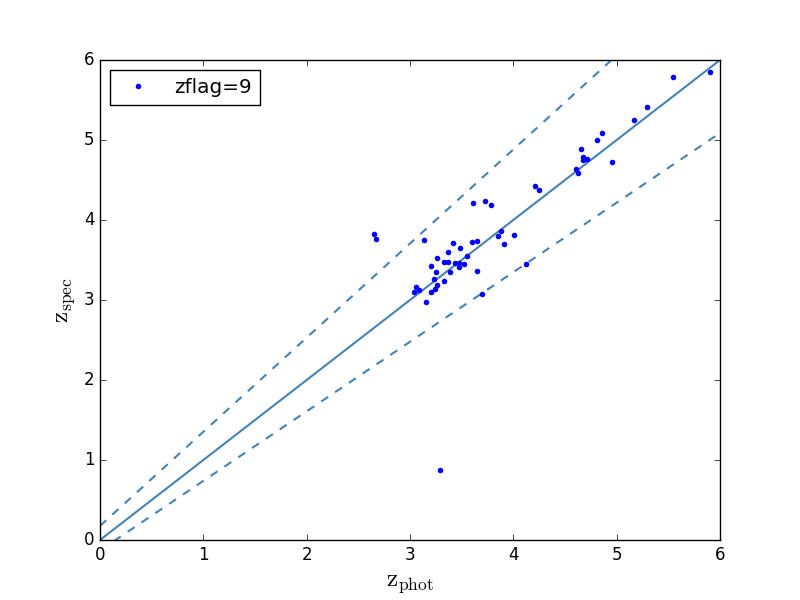}
\includegraphics[width = 10cm,clip=]{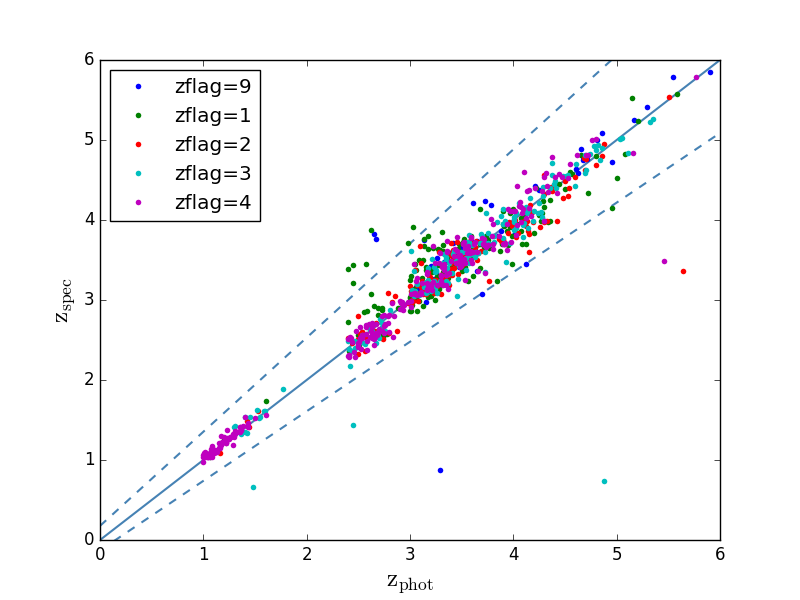}
\caption{The zphot vs zspec for all galaxies in the DR1 separated by flag type. The galaxies falling outside the dashed lines are the catastrophic outliers that have $|dz|>0.15$. }
\label{sepaflag}
\end{figure*}

\begin{figure*}
\includegraphics[width = 10cm,clip=]{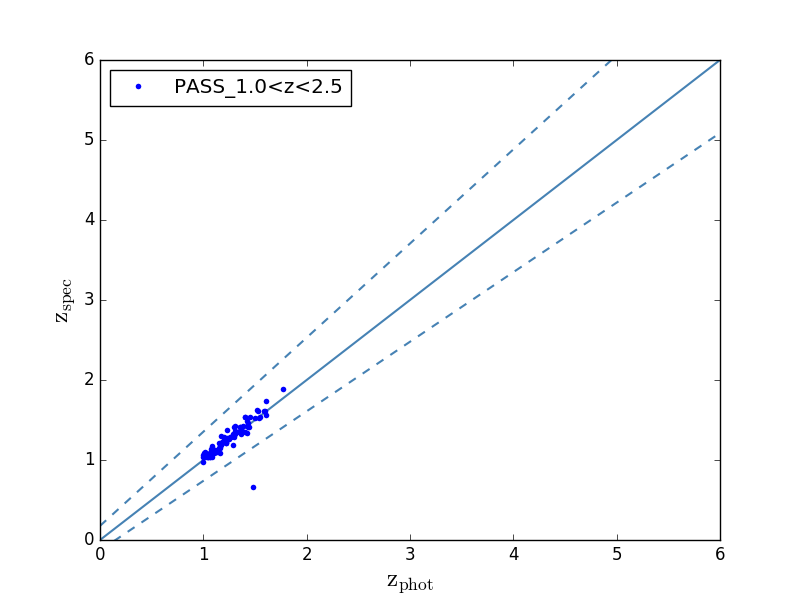}
\includegraphics[width = 10cm,clip=]{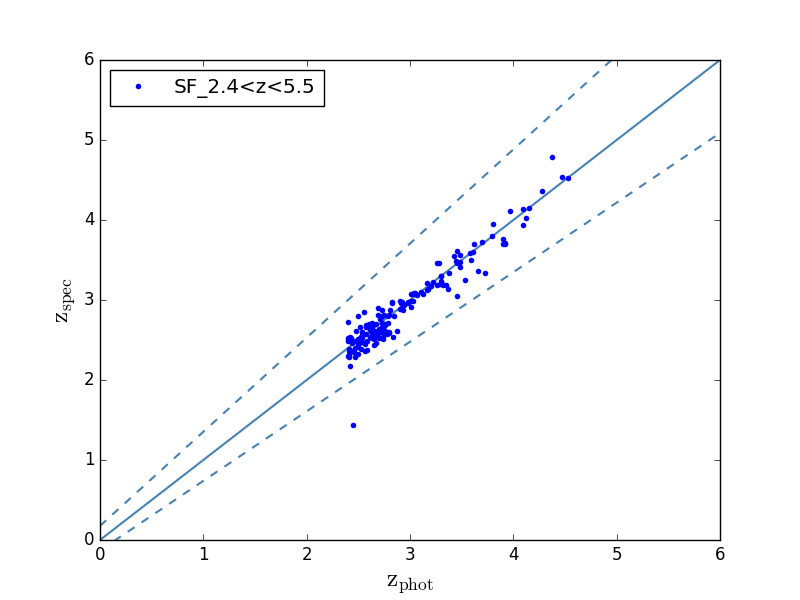}
\includegraphics[width = 10cm,clip=]{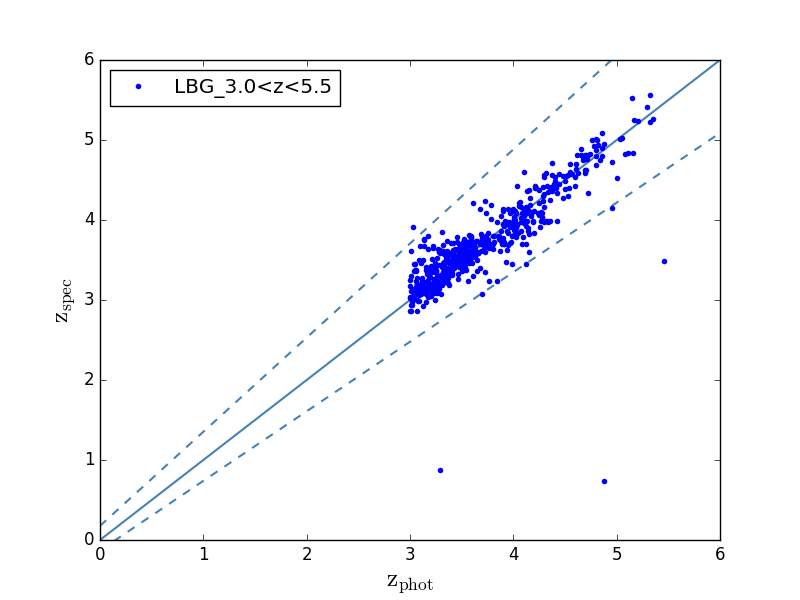}
\includegraphics[width = 10cm,clip=]{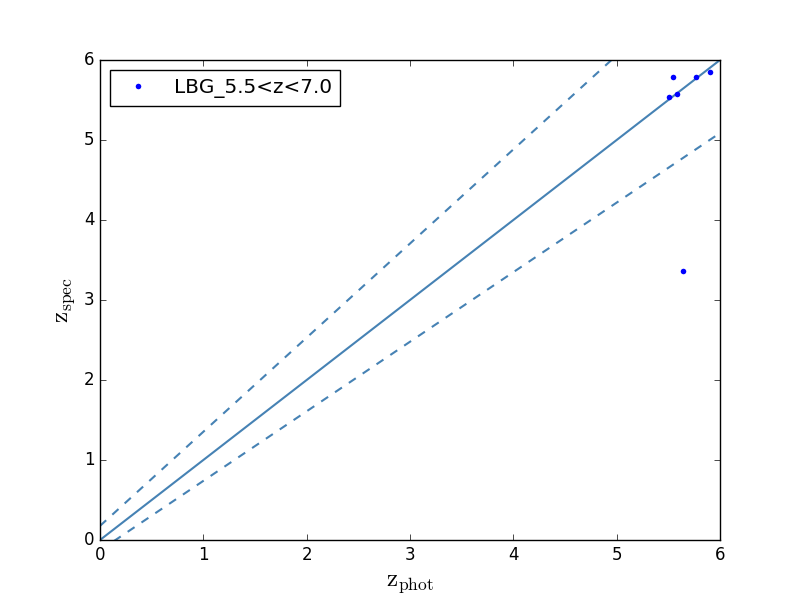}
\includegraphics[width = 10cm,clip=]{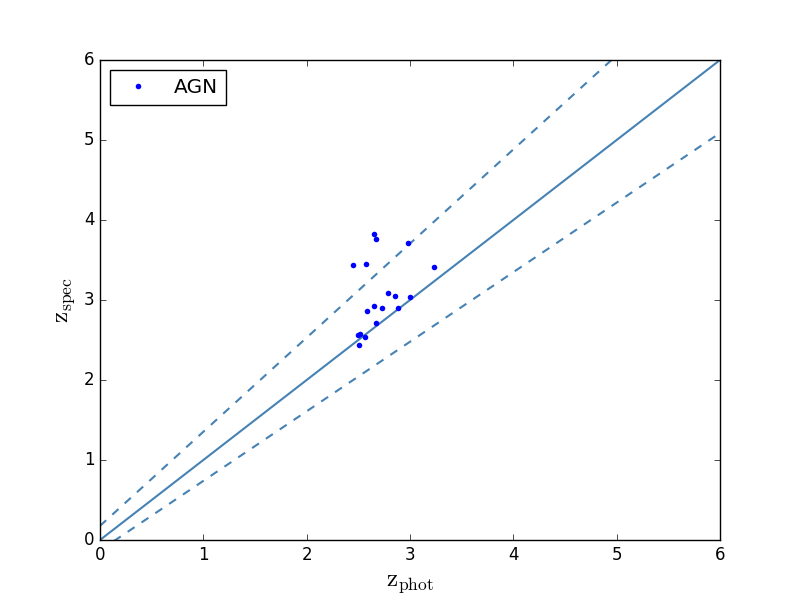}
\includegraphics[width = 10cm,clip=]{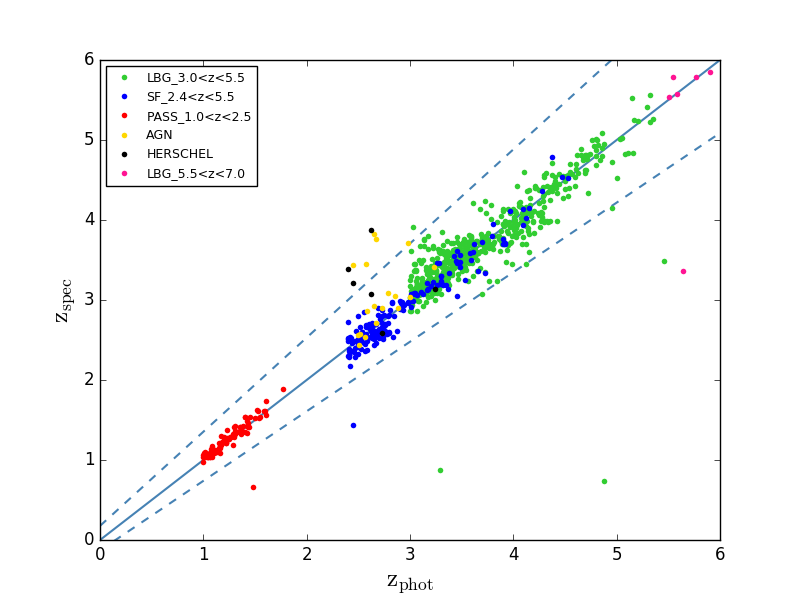}
\caption{The zphot vs zspec for all galaxies separated  by selection class type. The galaxies falling outside the dashed lines are the catastrophic outliers that have $|dz|>0.15$. }
\label{sepclass}
\end{figure*}

\begin{figure*}
\includegraphics[width = 10cm,clip=]{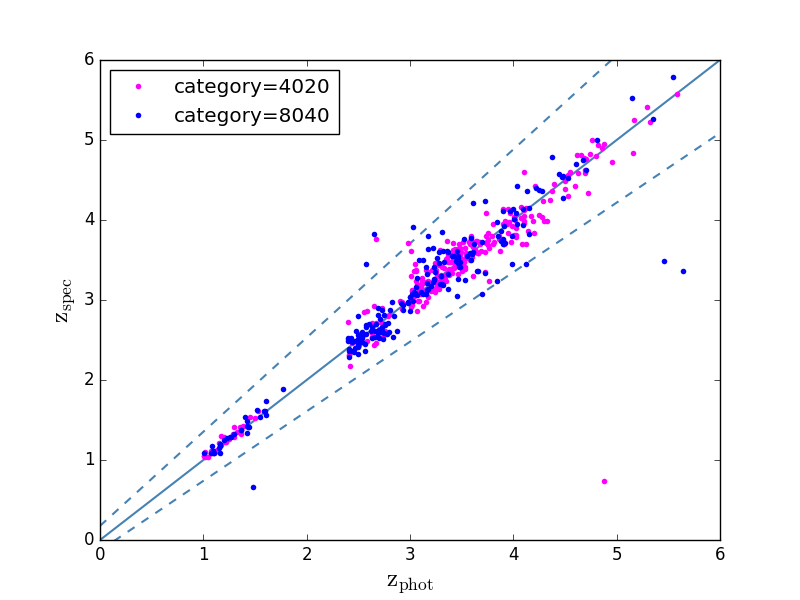}
\includegraphics[width = 10cm,clip=]{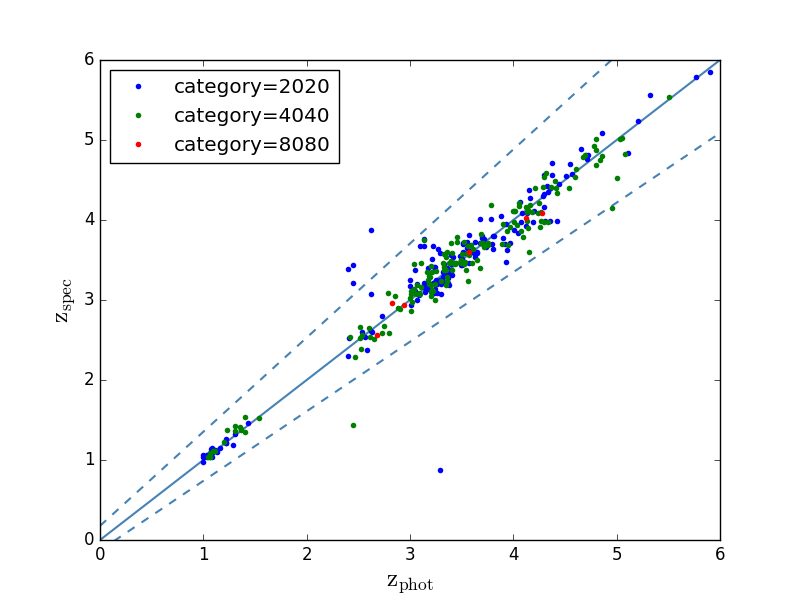}
\caption{The zphot vs zspec for all galaxies color-coded by  integration time: left are the galaxies which received half of the final integration time, and right the galaxies which received the full (20 , 40 or 80 hours) integration time.The galaxies falling outside the dashed lines are the catastrophic outliers that have $|dz|>0.15$. }
\label{septime}
\end{figure*}

In Figures \ref{sepaflag} and \ref{sepclass} we show the plots of the photometric redshifts vs spectroscopic redshifts for all galaxies that are part of the DR1 set, for which a redshift could be determined (873 objects), separated by flag type and target category. We also show in Figure \ref{septime} the same plot for objects that received the full final integration time (20, 40 or 80 hours) and those that received only half of the final allocation.
In Table \ref{tab:sigma} we report the number of outliers and the resulting scatter between the photometric redshifts and the spectroscopic redshifts, where the photometric redshift is the CANDELS one for the CANDELS footprint areas,  and the VANDELS one (McLure et al. 2018) for the outer areas.
This Table presents both the  the full scatter
 $\sigma_F =rms(\Delta z)$ where $\Delta z=(z_{spec}-z_{phot})/(1+z_{spec})$, and $\sigma_O$ derived after exluding the catastrophic outliers. We define a catastrophic outlier as a galaxy for which $|\Delta z|>0.15$.
We remark that $\sigma_O$  gives a non-optimal representation of the scatter since a few objects (i.e., the  outliers) can drive the scatter to large values. 
We also quantify  any systematic  bias  between photometric and spectroscopic redshifts, by $b=mean(\Delta z)$ after excluding the outliers. From the Table we see that the bias is much smaller than  the scatter.

In total there are  18 outliers, thus the resulting rate is  2.1\% which is in perfect agreement with the rate estimated from the accuracy of the VANDELS photometric redshifts, even if these were actually validated on a  sample with brighter average magnitude (see McLure et al. 2018 for details).
The number of outliers is higher for Flag 1 and 9 objects as expected, while it is extremely low for Flags 2,3,4. This  indicates that our flag system is actually conservative and probably the flags underestimate the reliability of the redshift. We notice that  8 of the 18 outliers are actually AGN or HERSCHEL selected sources.
In particular, for AGN the disagreement between photometric and spectroscopic redshift might be partly due to the fact that only templates of normal galaxies where used to determine the photometric redshifts \citep{salvato11,salvato09}.

\begin{figure}
\includegraphics[width =9.3cm,clip=]{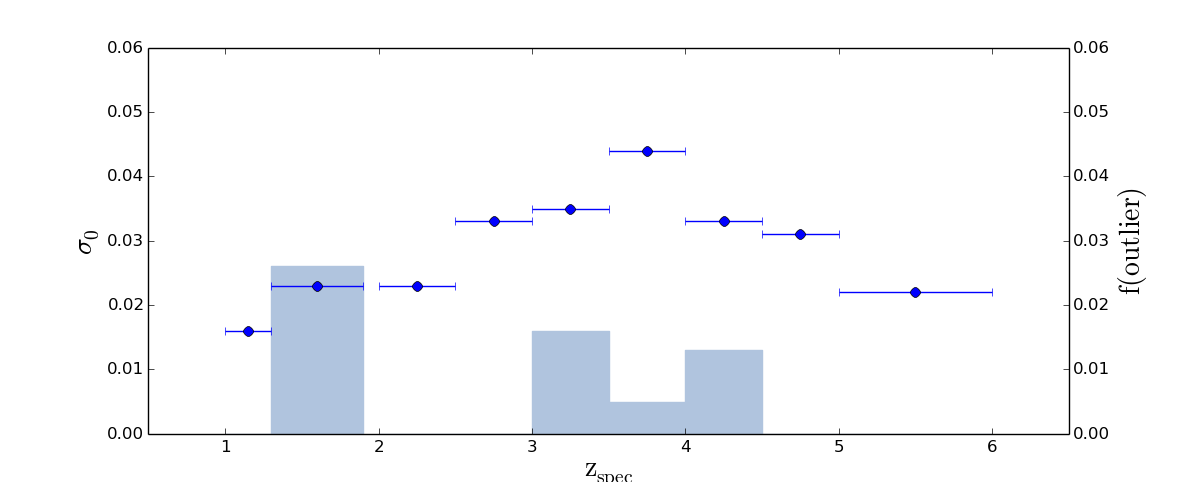}
\caption{Redshift dependence of the photometric redshift scatter and outlier fraction when comparing the  photometric redshift with the spectroscopic redshift. The blue dots show the scatter $\sigma_O$ (scaling on left-hand y-axis). The histograms show the fraction of outliers (scaling on right-hand y-axis). The sample does not include HERSCHEL and AGN selected targets.
}
\label{fig:sigma}
\end{figure}

If we further restrict the sample to the three main categories of targets of the VANDELS proposal, the outliers rate is actually just 1.2\% which is extremely low.
The Table shows that the agreement between zphot and zspec is excellent, for all quality flags with $\sigma_O$ clearly improving for higher flags. 
The class of targets with the highest overall accuracy is   the PASS one: this is easily explained both by the relative brightess of this sample (the median H-band magnitude is  $H=21.4$) and by the fact that the 4000 \AA\ break for galaxies in this redshift range  is very clear in the SED. 
In the same Table we report the  accuracy of the photometric redshifts divided by H-magnitude bins, after removing the HERSCHEL and AGN targets to have an unbiased view of the three main VANDELS groups. We find that  $\sigma_o$  decreases mildly as a  function of the H-band magnitude, while  the percentage  of outliers  increasing slightly at fainter magnitudes. This was also found by \cite{dahlenetal2013} for the CANDELS photometric redshifts, although in that case the fraction of outliers at magnitude fainter than 25 was much higher, $>10\%$ compared to our 2.3\% even in the faintest bin. 
We also investigate the  photometric redshift accuracy as a function of redshift, again after taking out the HERSCHEL and AGN targets.  Figure \ref{fig:sigma}  indicates that there is a  redshift trend in the photometric redshift accuracy, since the scatter $\sigma_o$  increases very smoothly from $\sigma_o =0.016$ at z$\sim 1-1.3$ to $\sigma_o =0.044$ at z$\sim 3.5-4$ and then decreases beyond that redshift. 

We  finally   compare the accuracy of photometric redshifts in the CANDELS footprint and in the outer areas.   The largest number of outliers is actually  found in the CANDELS areas (16) but this is due mostly  to the fact that all the Herschel targets and almost all of the AGN (i.e., the classes with the majority of interlopers) are selected in the CANDELS areas, and partly to the fact that the CANDELS targets are on average 0.4 magnitudes fainter than those in the wider areas. However the new VANDELS redshifts (i.e. those in the outer areas) also have a smaller bias and scatter. A possible explanation for the increased success rate  of the VANDELS photometric redshifts is that  all potential  targets were visually inspected before the final assembly of the catalogue, to reject obviously spurious and artefacts.

\begin{table*}
\begin{center}
\caption{Accuracy of photometric redshifts}
\begin{tabular}{lccccccccccccc}
\hline
Sample   &  $bias_z$ & $\sigma_F$ & $\sigma_O$  &  \%out & $Ntot$ &$Nout$   
        \\
\hline
All Flags & -0.002 &    0.106 &      0.036 & 2.1 &   873 & 18\\
Flag 1   &   0.005 &    0.061 &      0.047& 4.4 &   180 & 8 \\
Flag 2   &  -0.004 &    0.054 &      0.035 & 0.6 &   156 & 1 \\
Flag 3   &   0.001 &    0.184 &      0.031& 1.6 &   184 & 3 \\
Flag 4   &   0.005 &    0.039 &      0.029 & 0.3 &   295 & 1 \\
Flag 14  &   0.039 &    0.021 &      0.021  &  0.0 &     4 & 0 \\  
Flag 9   &   0.016 &    0.187 &      0.039 & 9.3 &   54 & 5 \\
\hline 
SFG 2.4<$z_{phot}$<5.5   &-0.008  &  0.044 &   0.032 &   0.6 &    177 & 1 \\
PASS 1.0<$z_{phot}$<2.5 &0.011  &  0.053 &   0.019 &  0.9 &    106 & 1 \\
LBG 3.0<$z_{zphot}$<5.5  &0.004 &  0.123 &   0.038 &  1.3  & 560 & 7 \\   
LBG 5.5<$z_{phot}$<7.0  &0.007 & 0.198 &   0.015 &  16.7 &  6 &1 \\      
AGN            &0.029  &  0.086 &   0.030 &  27.8 &  18 & 5 \\ 
HERSCHEL      & 0.059  &  0.116 &   0.069 &  50.0 & 6 & 3 \\
\hline
CANDELS footprint      &0.002 & 0.132  &  0.040      & 3.6  & 439  & 16 \\
Wide areas    &0.004  & 0.070  &  0.032  & 0.5  & 434 & 2  \\
\hline
H<21       & 0.008 & 0.018 &0.018 & 0  & 41 & 0 \\
H=[21-22]  & 0.012 & 0.074 & 0.019 & 2.0 &50 & 1\\
H=[22-23]  & -0.002 & 0.033 & 0.033 & 0.0 & 45 &0 \\

H=[23-24]  & -0.005 & 0.123 & 0.037 & 1.6 & 129 &2 \\
H=[24-25]  & 0.001 & 0.165 & 0.038 &  0.9 & 222  &2  \\
H=[25-26]  & 0.001 & 0.047 & 0.035 &  1.2 & 254  &3  \\
H>26         & 0.014 & 0.073 & 0.042 & 2.3 & 87  &2 \\
\hline\hline
\end{tabular}
\label{tab:sigma}
\end{center}
\end{table*}

\subsection{Examples of data products}
\begin{figure}
\includegraphics[width = 10cm,clip=]{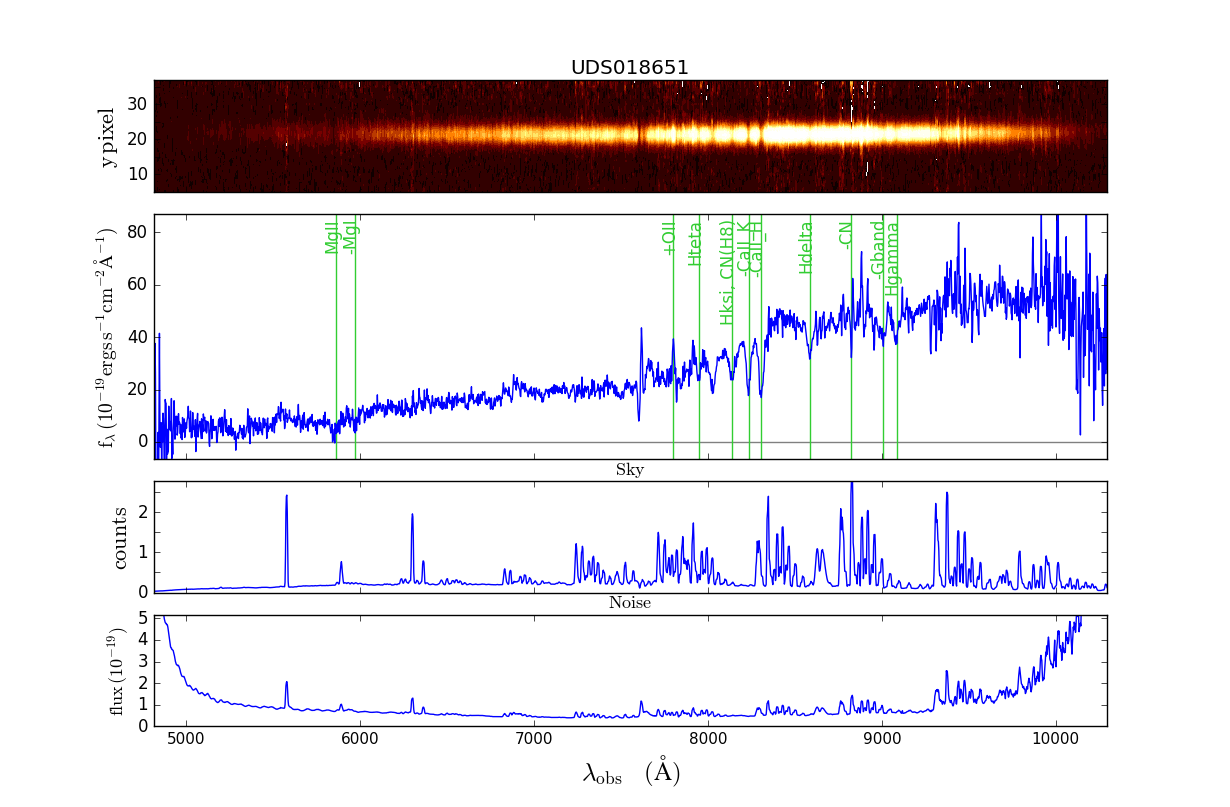}
\caption{From top to bottom , the 2-dimensional spectrum, 1-dimensional extracted spectrum,  sky counts and rms noise of target VANDELS-UDS-018651. This object was selected as a PASS galaxy and has a redshift 1.093 with Flag 4. The main spectral features identified  are indicated in the 1-D spectrum.}
\label{obj1}
\end{figure}

\begin{figure}
\includegraphics[width = 10cm,clip=]{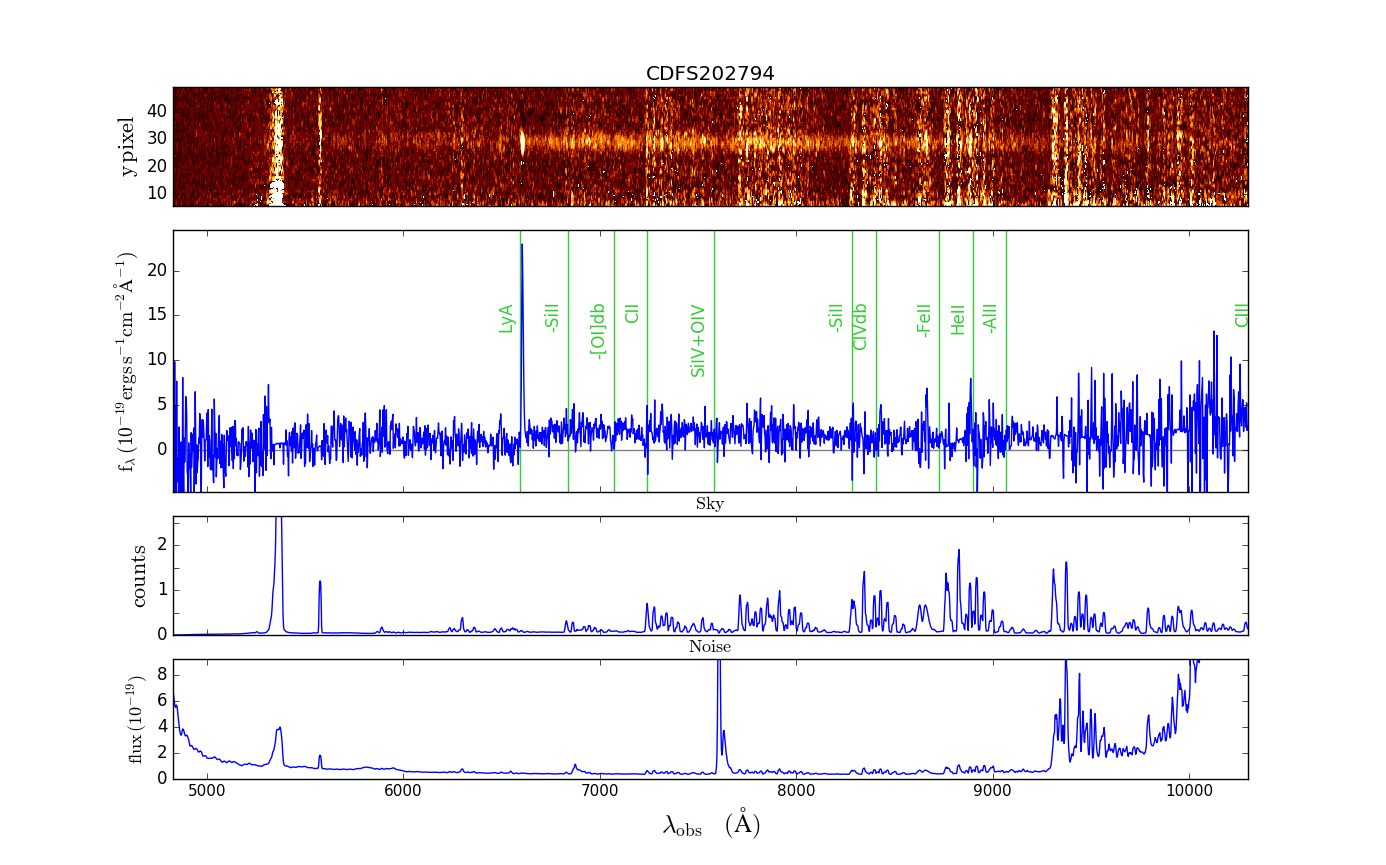}
\caption{From top to bottom, the 2-dimensional spectrum, 1-dimensional extracted spectrum,  sky counts and rms noise of target VANDELS-CDFS-202794. This object has an H-band magnitude of 24.1 and was selected in the category LBG. It was assigned a $z_{spec}=4.4266$ and Flag 4. The main spectral features identified (emission lines and interstellar absorption lines)  are indicated in the 1-D spectrum.}
\label{obj2}
\end{figure}

\begin{figure}
\includegraphics[width = 10cm,clip=]{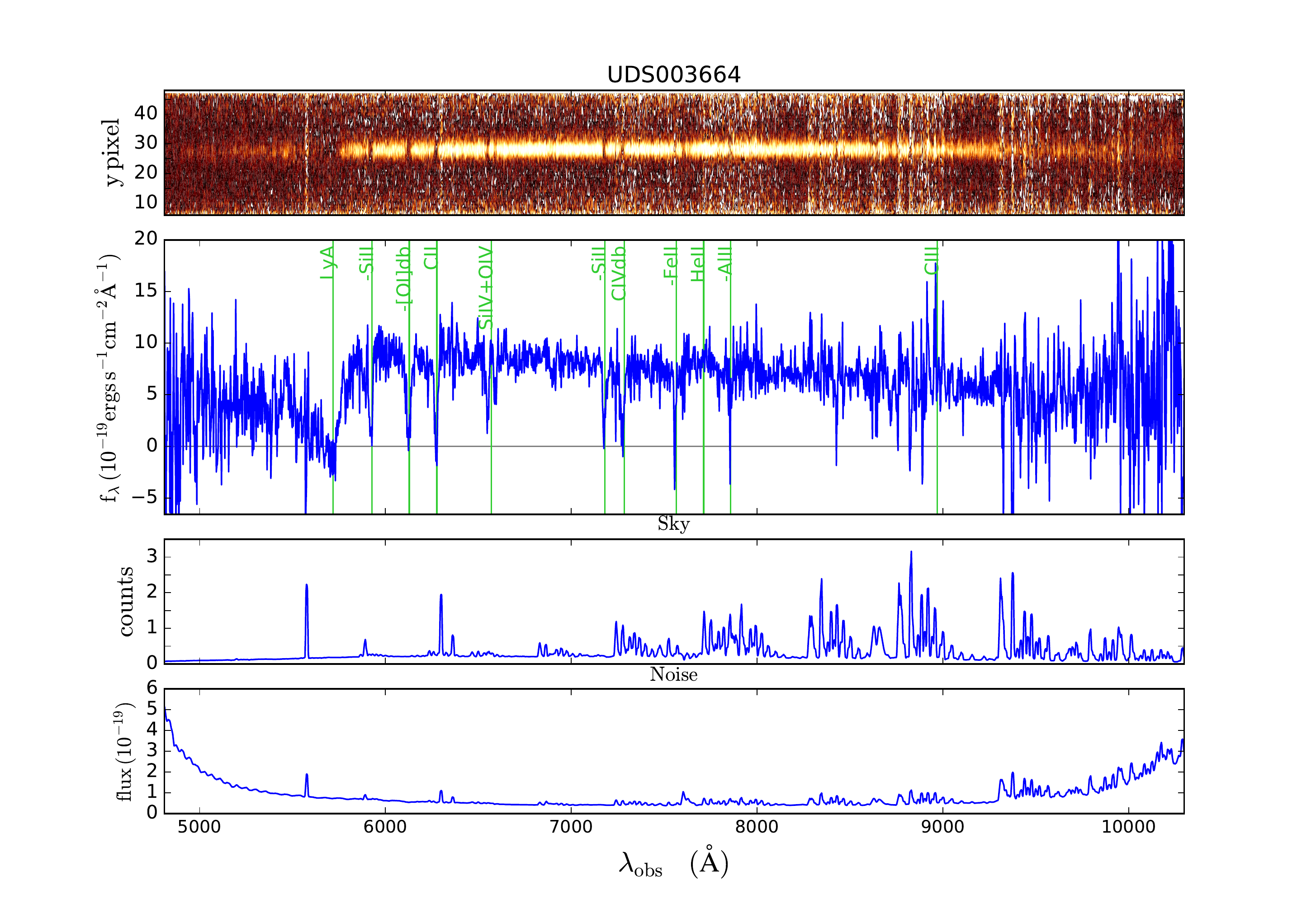}
\caption{From top to bottom, the 2-dimensional spectrum, 1-dimensional extracted spectrum,  sky counts and rms noise of target VANDELS-UDS-003664. This object has an H-band magnitude of 23.1 and was selected in the category SFG. It was assigned a $z_{spec}=3.703$ and Flag 4. The main spectral features identified (the Ly$\alpha$ absorption  and several interstellar absorption lines)  are indicated in the 1-D spectrum.}
\label{obj3}
\end{figure}

\begin{figure}
\includegraphics[width = 10cm,clip=]{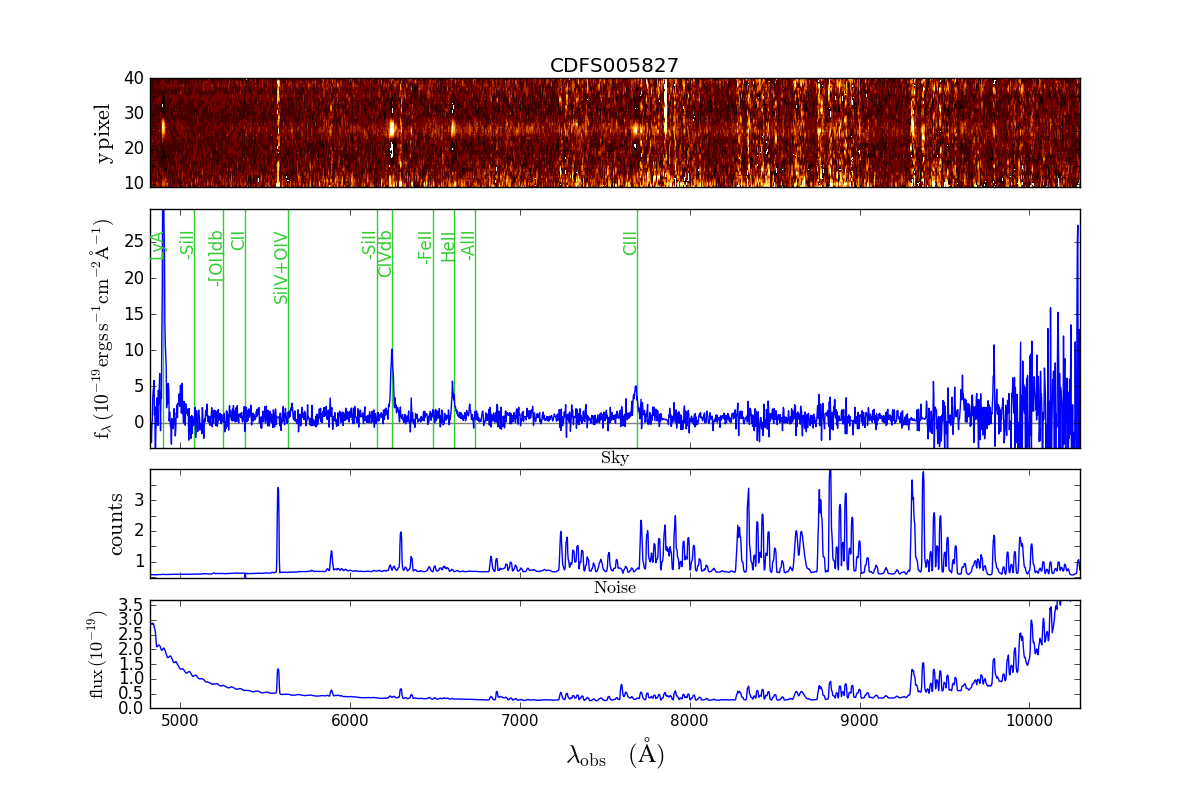}
\caption{From top to bottom, the 2-dimensional spectrum, 1-dimensional extracted spectrum,  sky counts and rms noise of target VANDELS-CDFS-005827. This object was selected as an AGN target and indeed it is identified as an AGN (quality flag 14) at z=3.0328.  The main spectral features identified  (the Ly$\alpha$, CIV, HeII and CIII] emission lines)  are indicated in the 1-D spectrum.}
\label{obj4}
\end{figure}

\begin{figure}
\includegraphics[width = 10cm,clip=]{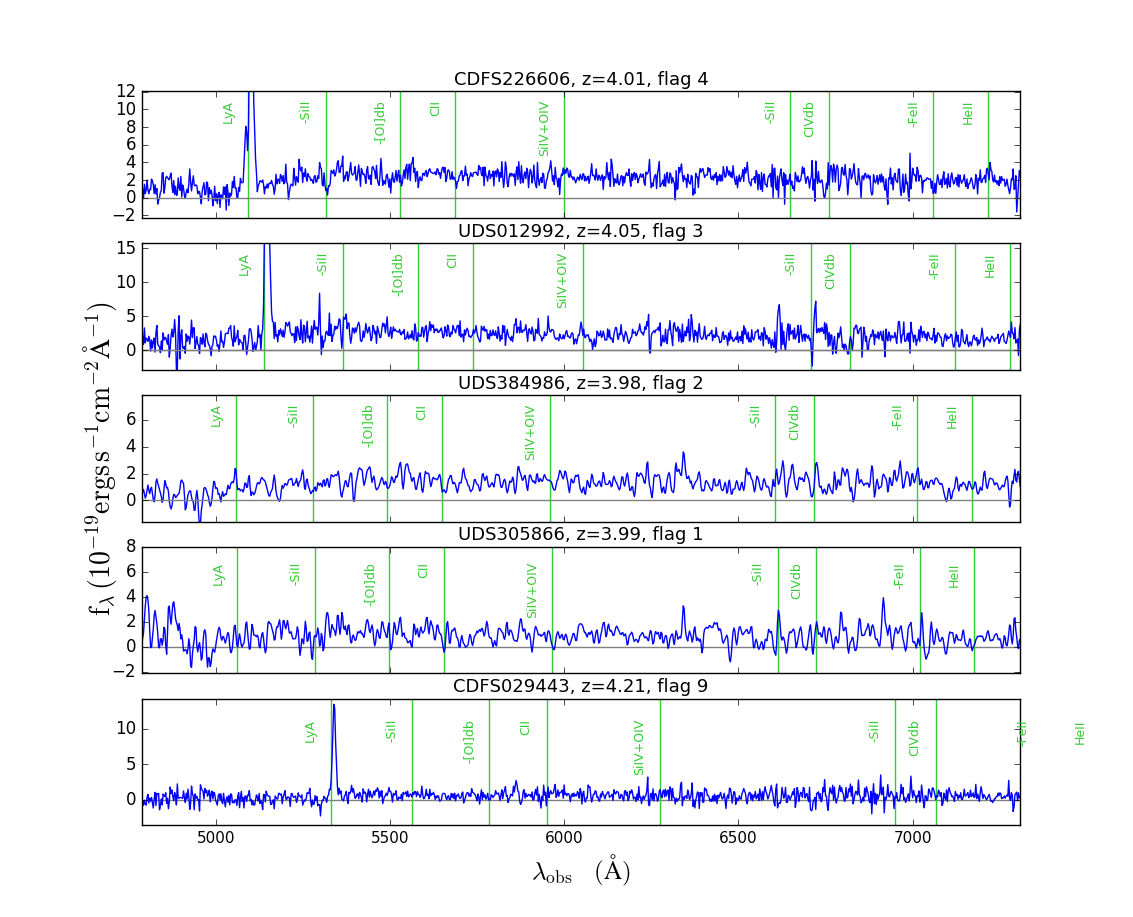}
\caption{From top to bottom, 1-dimensional extracted spectra of five VANDELS  LBGs at redshift z$\sim 4$ with different quality flag:  VANDES-CDFS-226606 with QF=4, VANDELS-UDS-012992 with QF=3, VANDELS-UDS-384986 with QF=2, VANDELS-UDS-305866 with QF=1 and VANDELS-CDFS-029443 with QF=9. The spectra of the QF=1 and 2 galaxies have been slightly smoothed to better show the absorption  lines. }
\label{fig:flag}
\end{figure}

To illustrate the content of the VANDELS-DR1, we present some examples of data products for different types of targets and different quality flag. In Figures \ref{obj1},\ref{obj2},\ref{obj3},\ref{obj4} we show the 2-dimensional, 1-dimensional, sky and  noise spectra of four targets, respectively selected as PASS, LBG, SFG and AGN. All these galaxies have quality flags 4, since the redshift is determined with great accuracy from the many emission and or/absorption lines visible in the spectra.
To illustrate the meaning of the different quality flags, in Figure \ref{fig:flag} we present the spectra of five galaxies, all  selected as LBGs, and approximately at the same redshift ($z\sim 4$),  but with quality flags 4, 3, 2, 1 and 9 respectively. In the Flag 4 spectrum we can clearly identify two emission lines (the Ly$\alpha$ and HeII lines) and several absorption lines such as the SiII, OI, CII, CIV etc. The drop in the continuum flux below the Ly$\alpha$ line is also very clear.
In the Flag 3 spectrum we identify the Ly$\alpha$, several absorption lines (e.g. the SiII and CIV)  and the drop in the continuum flux. In the Flag 2 spectrum  (which has been smoothed in the Figure for clarity), we identify the SiII, CII and SiIV lines with good confidence and the drop in the continuum flux is also clear. In  the Flag 1 only the SiIV line is clearly identified, but there is a general good agreement when cross correlating this spectrum with templates at the assigned redshift. Finally in the  Flag 9  case, there is only one bright emission line that is detected in the spectrum.  The line does not show a prominent asymmetry (which would clearly identify it as Ly$\alpha$) and no continuum is detected blueward of the line, therefore there remains  some ambiguity in the redshift identification.
\section{Summary}
We have presented the first Data Release of the VANDELS public spectroscopic survey. VANDELS, a deep survey of the CANDELS CDFS and UDS fields is an ESO  public survey carried out with VIMOS and has  obtained more than 2000 ultradeep medium resolution spectra of galaxies in the wavelength range 4800-10000 \AA.
The  DR1 is the release of all  spectra obtained during the first season of observations, and it includes all targets for which either the scheduled  integration time  or half of the total time was completed. The release includes spectra for 879 objects, 464 in the UDS and 415 in the CDFS.  Together with the spectra, we release also the spectroscopic redshifts measured  by the collaboration, with a quality flag that assesses their reliability. In this paper we present the  statistics of the redshift quality and dicuss the excellent accuracy of the VANDELS photometric redshifts, with an outlier rate as low as  2.1\% and overall accuracy of $\sigma_o=0.035$, which improve if we restrict the statistics to the main VANDELS target categories.  We also present some examples of data products, to illustrate the content of the release. All spectra and information are available in the ESO archive.
The second Data Release will include all spectra completed during the second observational season, and   will be available at the end of  Summer 2018. A final release is expected for June 2020 and  will include 
improved re-reduction of the entire spectroscopic data,  a series of galaxies' physical properties (stellar masses, SFRs, dust attenuation  etc)  derived by the collaboration through SED fitting, and measurements of absorption and emission lines identified in the VANDELS spectra. 

\section*{Acknowledgement}
Based on data products from observations made with ESO Telescopes at the La Silla Paranal Observatory under programme ID 194.A-2003(E-K). We thank the ESO staff
for their continuous support for the
VANDELS  survey, particularly the Paranal staff, who helped us to conduct the observations, and the  ESO user support group in Garching. RJM, AM, EMQ and DJM acknowledge funding from the European Research Council, via the award Consolidator Grant (P.I. R. McLure)
AC aknowledges the grant PRIN-MIUR 2015 and ASI n.I/023/12/0. PC acknowledges support from CONICYT through the project FONDECYT regular 1150216. RA acknowledges support by the ERC Advanced Grant 695671 “QUENCH”. FB  acknowledges the support by Funda\c{c}\~ao para a
Ci\^encia e a Tecnologia (FCT) via the postdoctoral fellowship
SFRH/BPD/103958/2014 and through the research grant
UID/FIS/04434/2013.

\bibliographystyle{aa}
\bibliography{ref.bib}

\end{document}